\documentclass[graybox]{svmult}

\usepackage{type1cm}        
\usepackage{makeidx}         
\usepackage{graphicx}        
\usepackage{multicol}        
\usepackage[bottom]{footmisc}
\usepackage{url}

\usepackage{newtxtext}       %
\usepackage{soul}            
\usepackage{newtxmath}       
\usepackage{tabularx}
\usepackage{orcidlink}
\usepackage{cite}

\makeindex             

\newcommand{\etal}{\textit{et al.}}
\newcommand{\bcso}{BaCuSi$_2$O$_6$}
\newcommand{\uno}{$H_{c1}$}
\newcommand{\dos}{$H_{c2}$}

\begin{document}

\title*{Bose-Einstein Condensation of Magnons in \bcso: An experimental perspective}
\author{Marcelo Jaime$^{[0000-0001-5360-5220]}$ and Franziska Weickert$^{[0000-0002-1545-9645]}$}
\institute{Marcelo Jaime \at Physikalisch-Technische Bundesanstalt, Bundesalle 100, Braunschweig 38116, Germany \email{marcelo.jaime@ptb.de}
\and Franziska Weickert \at Physikalisch-Technische Bundesanstalt, Bundesalle 100, Braunschweig 38116, Germany \email{franziska.weickert@ptb.de}}
%
%
\maketitle

\abstract{Han Purple, a pigment first obtained in ancient China, is one of the earliest known synthetic pigments. Also naturally occurring, as a mineral, it is known as colinowensite (Cwn). Its chemical formula is \bcso, and its structure is a layered cyclosilicate in which magnetic Cu$^{2+}$ ions ($S = 1/2$) form dimers arranged on a square lattice, making it also the first known synthetic metal dimer compound. Most interesting magnetic properties arise from a strong intradimer spin coupling, accompanied by weaker interdimer interactions within and between Cu-dimer layers. In zero or small magnetic fields, \bcso\ remains a quantum paramagnet. However, under high magnetic fields between 23 and 49 Tesla  \textemdash about a million times stronger than Earth's magnetic field\textemdash~ it undergoes magnetic ordering at sub-liquid Helium temperatures into an almost ideal easy-plane ($XY$) antiferromagnetic state regarded as a realization of a Bose-Einstein condensate of magnons. Within this experimentally accessible field range, \bcso\ serves as an extraordinary playground for testing predictions of quantum many-body physics.}

\section{Background}
\label{sec:1}

A nearly three-thousand-year-long human fascination with silicates (i.e., note that silica is the raw material used in the semiconductor industry) likely began with the ability to create pigmented glass beads —essentially out of earth and fire— that could shine in unearthly shades of blue. Inclusions of cyclosilicates containing barium and pairs of copper atoms can shine purple, a non-spectral color,\footnote{Purple, different from single-wavelength violet, is created by the human brain combining blue and red lights. A perceptual color, not visible in the rainbow,  results from an absorption band that extends from $\sim$490 nm to $\sim$600 nm effectively filtering out the green in the pigment crystals.} which has consistently been regarded as a symbol of grandeur and holiness, representing concepts of monarchy, dignity, and wealth.

Han Purple, chemically known as \bcso, is a fascinating material that bridges ancient craftsmanship and quantum physics. It was first synthesized in ancient China during the Zhou Dynasty (1045$-$256 BC), identified in a rhombic glassy faience bead dated to 777$-$766 BC \cite{ma2006, berke2006}, and it saw extensive use in the Han Dynasty (206 BC$-$220 AD), from which it derives its modern name \cite{fitzhugh1992}. Unlike naturally occurring pigments such as ocher or malachite, Han Purple was a technological innovation: a fully synthetic pigment created through sequential high-temperature solid-state reactions, requiring controlled heating to approximately 800$-$900$^o$C \cite{qin2016}. While lower temperature synthesis was recently achieved  using a  hydrothermal reaction method \cite{cormar2019} inside a high-pressure sealed autoclave container, the technical prowess needed is no simpler to achieve. These processes demonstrate an exceptional level of materials understanding for the time, involving the complex manipulation of locally available minerals containing barium, copper, and silicon.

Archaeological discoveries have found Han Purple in tomb murals, ceremonial vessels, the Terracotta Army, and pottery, often alongside Han Blue (BaCuSi$_4$O$_{10}$), a closely related blue pigment \cite{berke2002, berke2006}. Han Purple’s vivid, deep hue symbolized wealth, status, and perhaps even spiritual power, and its presence in royal tombs suggests it was highly prized. Despite their chemical similarity, Han Purple is structurally less stable than Han Blue, contributing to its relative scarcity in surviving artifacts. The production of Han Purple may have originated from early Taoist monks’ efforts to create synthetic jade or glass, both of which were highly valued in ancient Chinese culture. The complexity of its synthesis implies that early Chinese chemists had practical knowledge of mineral mixing, firing temperatures, and phase transformations—even if the theoretical underpinnings were unknown to them. With the decline of Taoism, the knowledge and interest in Han Purple and similar pigments were lost, until their recent resurfacing following the discovery of Emperor Qin Shi Huang’s Terracotta Warriors on 29 March 1974, when farmer Yang Zhifa uncovered fragments of pottery while digging a well near Xi'an, China \cite{wikipedia}. Notably, \bcso\ also exists in natural form, named colinowensite (Cwn) after its discovery in the Kalahari Manganese Field, South Africa, by rock collector Colin R. Owen \cite{rieck2015}.

The rediscovery of Han Purple due to its fascinating magnetic properties in modern times \cite{finger1989, sasago1997, sparta2004} has opened a second chapter. While ancient pigments capture the imagination of art historians \cite{bouherour2001, berke2006}, and the trained eyes of mineral diggers \cite{rieck2015}, \bcso\ also captivates condensed matter physicists due to its unique quantum magnetic properties. The very same copper ions (Cu$^{2+}$) that connect silicate rings to yield the pigment’s color \cite{garcia2015} also form a two-dimensional array of quantum spins $S = 1/2$, arranged in Cu$^{2+}$-Cu$^{2+}$ dimers within the crystal lattice. In the absence of an external magnetic field, these dimers form an energy-gap quantum paramagnetic ground state . When subjected to very high magnetic fields, the energy gap closes by the Zeeman effect, and a region opens where the system exhibits a field-induced quantum phase transition (QPT) into a distinct magnetic order \cite{jaime2004, sebastian2006a, allenspach2022}. The kind of magnetic order has been a topic of intensive scientific discussions, because universal physical concepts like Bose-Einstein Condensation (BEC) of integer spin particles can be used to model \bcso, and to draw specific predictions about its thermodynamic and transport properties that - in turn - can be tested experimentally.

Han Purple is more than a historical curiosity; it is a profound reminder that ancient materials science can unexpectedly intersect with the forefront of modern physics. From decorating the tombs of emperors to testing theories of quantum criticality in reduced dimensions in today’s laboratories, \bcso\ embodies a rare continuum of human ingenuity across millennia. In this chapter, we review the intense condensed matter physics research on \bcso\ and closely related compounds over the last 20 years and provide a future outlook.

\section{Synthesis}
\label{sec:2}

\subsection{Solid state reaction} 
The synthesis of \bcso\ in powder (polycrystal) form can be achieved via a two-step high-temperature solid-state reaction using high-purity barium carbonate (BaCO$_3$), copper(II) oxide (CuO), and silicon dioxide (SiO$_2$), usually in finely powdered form.

The powders are measured in a stoichiometric ratio: one part BaCO$_3$, two parts CuO, and one part SiO$_2$. These materials are then thoroughly mixed, either manually using a mortar and pestle or mechanically in a ball mill, to ensure a completely uniform distribution. The mixed powder is placed in an alumina crucible and heated gradually to about 800$-$850°C in air. This first heating stage, known as calcination, serves two purposes: it decomposes the barium carbonate into barium oxide and carbon dioxide, and it initiates the chemical reactions between the barium oxide, copper oxide, and silica.

After maintaining the temperature for 12 to 24 hours, the material is cooled to room temperature. Afterwards, the partially reacted material is ground again to break up lumps and to remix the components thoroughly. Next, the powder is subjected to a second temperature firing at higher temperatures, typically around 950$-$1000°C. This step drives the reaction to completion and forms the final \bcso\ compound. The material is kept at this high temperature for another 12 to 24 hours to ensure full crystallization. Once the reaction is complete, the sample is cooled slowly to room temperature to avoid introducing cracks or unwanted secondary phases. In some protocols, a further annealing step is performed at about 900°C after another round of grinding to improve the crystallinity and purity of the sample \cite{qin2016, finger1989, sebastian_thesis}.

The resulting polycrystalline material is usually analyzed by powder X-ray diffraction (XRD) to confirm that the desired BaCu$_{2}$SiO$_{6}$ phase has formed without significant impurities. Careful control of temperature, stoichiometry, and mixing is crucial, as common impurities can include Ba$_2$SiO$_4$, BaCuSi$_4$O$_{10}$ (Han Blue), or Cu$_2$O.

\subsection{Traveling molten floating zone}
To obtain single crystals (SCs) with the traveling molten zone technique,  dense polycrystalline \bcso\ is first prepared to form a solid, cohesive rod. The rod is then mounted vertically in a mirror furnace, which uses intense, focused light from halogen lamps and gold-plated mirrors to create a small, localized molten zone in the material. This (molten) high temperature zone is carefully moved — or "floated" — along the length of the rod from top to bottom by slowly translating either the rod, the light source, or both.

As the molten zone travels, the material re-solidifies behind it as a single crystal, growing with the structure seeded by the initial melt. The process requires precise control over the temperature gradient, the translation speed (usually a few mm per hour), and the rotation of the rod to stabilize the molten zone and encourage high crystal quality. In the case of \bcso, it is important to perform the growth in an oxygen atmosphere to avoid the formation of red cuprite (Cu$_2$O) inclusions. Careful X-ray diffraction must be used to characterize the crystals obtained thereafter, as the ingrowth of relatively large (several mm long) crystals is common.

Single crystal samples grown by this method were used, among others, by Sasago \textit{et al.}, Sparta \textit{et al.}, Jaime \textit{et al.}, R\"{u}egg \textit{et al.}, Kr\"{a}mer \textit{et al.}, and Puphal \textit{et al.} \cite{sasago1997, sparta2004, jaime2004, horvatic2005, zvyagin2006, sebastian_thesis, kramer2007, qin2016, vanwell2016, puphal2019}. A comprehensive step-by-step description of the material preparation steps and characterization can be found here \cite{sebastian_thesis}.

\subsection{Flux method}
In the flux method, a molten solvent (the "flux") is used to lower the effective melting point, enabling \bcso\ single crystals to grow slowly at reduced temperatures. Typical fluxes include mixtures of salts such as BaCl$_2$, BaO, or other compounds that are chemically compatible with \bcso.

The starting materials—BaCO$_3$, CuO, and SiO$_2$—are mixed with an excess amount of the chosen flux and placed in a crucible, commonly made of platinum or alumina. The mixture is then heated to a temperature sufficient to dissolve the reactants into the flux, typically several hundred degrees Celsius below the actual melting point of \bcso.

As the melt is slowly cooled, \bcso\ crystals gradually precipitate out. The cooling rate is carefully controlled, typically in the range of 1$-$5$^\circ$C per hour, to promote the growth of large, well-formed single crystals.

After crystal growth is complete, the residual flux is removed either mechanically or by dissolving it in a suitable solvent (e.g., water or acid, depending on the flux composition).

Single-crystal samples grown using this method were employed by Sebastian \textit{et al.}, Samulon \textit{et al.}, and Jaime \textit{et al.} \cite{sebastian2005, samulon2006, sebastian2006a, sebastian2006b, harrison2006, sebastian2007, puphal2019}. A comprehensive, step-by-step description of the material preparation and characterization procedures can be found here \cite{sebastian_thesis}. Puphal \textit{et al.} \cite{puphal2019} grew with this technique Sr-doped \bcso\ crystals, and achieved a simplified version of \bcso\ where a cryogenic temperature crystallographic $c$-axis modulation is absent, resulting in a cleaner magnetic ground state.

\subsection{Growth by hydrothermal reaction}

Crystal growth by hydrothermal reaction is a method used to grow polycrystalline samples from aqueous solutions at moderate temperatures and high pressures, often in sealed vessels. This technique mimics natural geological processes and is widely used to grow quartz, emeralds, and various synthetic materials, such as zeolites and complex oxides.

The synthesis of Han Purple (\bcso) pigment particles under hydrothermal conditions can be achieved at significantly lower temperatures, in the range of 250$^\circ$C to 300$^\circ$C. The precursor analytical-grade reagents are barium chloride (BaCl$_2\cdot$2H$_2$O, 99\% purity), copper nitrate hemihydrate (Cu(NO$_3$)$_2\cdot$2.5H$_2$O, 98\% purity), copper chloride (CuCl$_2$), and sodium metasilicate nonahydrate (Na$_2$SiO$_3\cdot$9H$_2$O, 98\% purity). Prior to hydrothermal treatment, the powdered reagents are weighed according to the stoichiometric molar ratio Ba:Cu:Si = 1:1:2, corresponding to the \bcso\ pigment composition.

Based on this ratio, the dry-mixed quantities are: BaCl$_2\cdot$2H$_2$O\,=\,0.2418\,±\,0.0005\,g, Cu(NO$_3$)$_2\cdot$2.5H$_2$O\,=\,0.2302\,±\,0.0005\,g, and Na$_2$SiO$_3\cdot$9H$_2$O = 0.5678 ± 0.0005 g. An alternative preparation method uses 0.1304 ± 0.0005\,g of CuCl$_2$. The powder mixture is placed at the bottom of a Teflon vessel and dissolved by adding 15\,ml of the selected hydrothermal medium—either alkaline NaOH solution or deionized water. This volume corresponds to a 50\% filling ratio of the inner volume of the micro-autoclave Teflon vessel.

Two distinct synthesis routes were explored by Corona-Martinez \etal \cite{cormar2019}: one using NaOH solutions with varying concentrations (1$-$4 M), and the other using deionized water. The dissolution of the solid mixture is carried out under magnetic stirring at 400\,rpm for 30\,minutes, promoting the formation of a light sky-blue coprecipitated gel precursor. The sealed autoclave is then placed in a convection oven and heated to the desired reaction temperature. Treatments are conducted at temperatures between 180$-$240$^\circ$C for durations ranging from 3 to 96\,hours.

The reaction products are recovered by gravimetric separation, then washed thoroughly four times with hot water ($60^\circ$C), and dried overnight in an oven at $80^\circ$C. Powder samples produced by this method were grown and characterized by Chen \textit{et al.}\cite{chen2014} and Corona-Martinez \textit{et al.}\cite{cormar2019}.

\begin{figure}[ht]
\includegraphics[scale=.55]{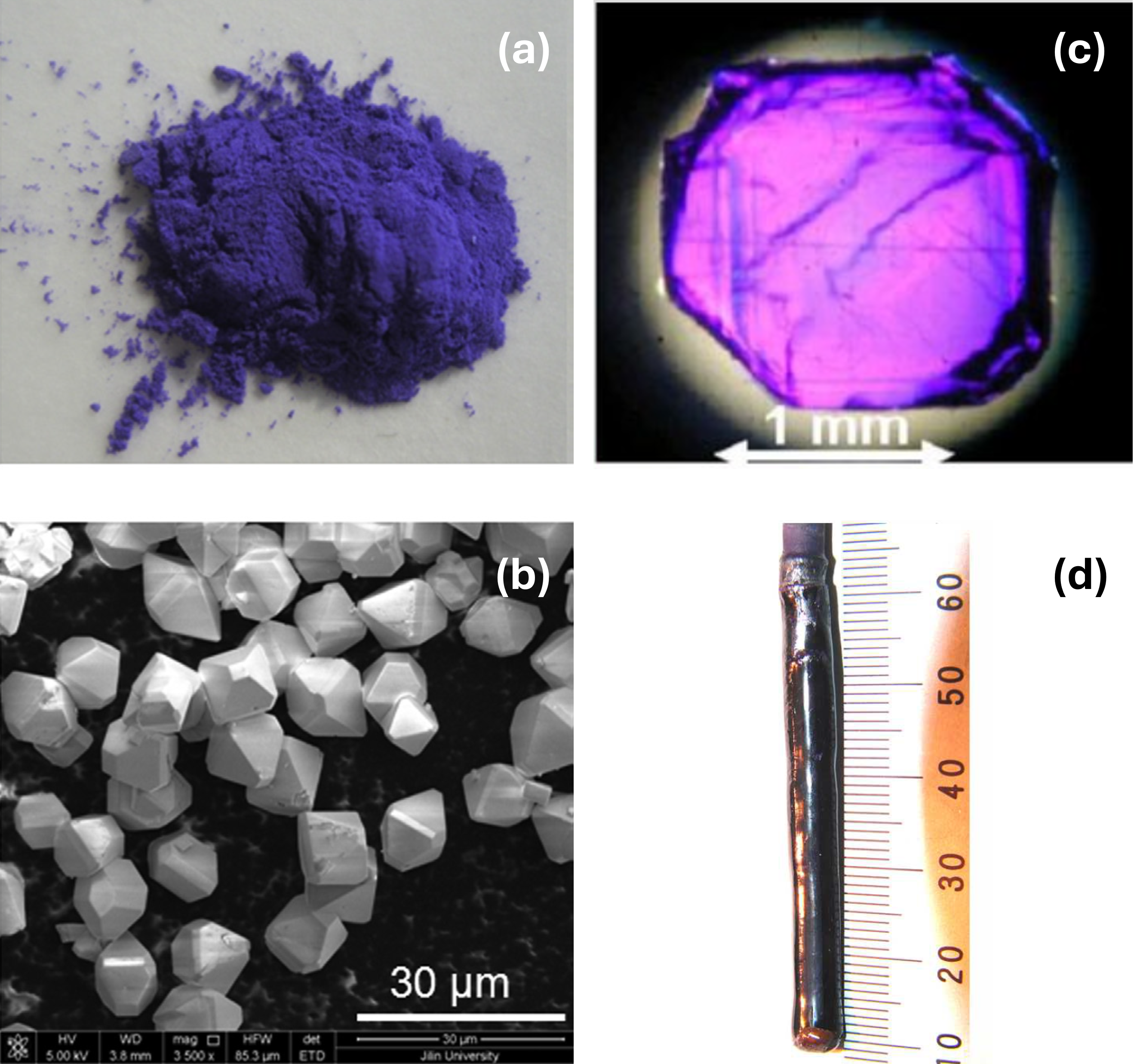}
\caption{(a) Han purple \bcso\ powder synthesized by the hydrothermal route at $250^o$C for 48\,hrs. (b) SEM image showing $10\mu$m big crystals. Reprinted from Chen \textit{et al.} \cite{chen2014} copyright {\copyright}2014 with permission from Elsevier. (c) Single crystal sample grown using the flux technique. Depending on the flux used, the crystal size range is 1\,-\,10\,mm. Reprinted from van Well {\it et al.}, \cite{vanwell2016} copyright {\copyright}2016 American Chemical Society. (d) 60\,mm long crystalline rod, often consisting of a few single crystals inter-grown together, obtained with the traveling molten floating zone method. Reprinted with permission from S.E. Sebastian, PhD. Thesis, \cite{sebastian_thesis}(Copyright {\copyright}2006, Suchitra E. Sebastian)}
\label{fig:1}  
\end{figure}

From the vast literature on growing single and polycrystalline samples \bcso\ by different methods, it cannot be concluded that there exists one specific method to achieve superior quality. Instead, it is suggested that high-quality samples can be obtained by rather different approaches, and the method should be chosen according to the planned experimental apparatus/technique constraints and requirements in the following investigations.

\section{Crystal structure}
Now that we have reviewed methods for growing Han Purple, we turn to the characterization techniques used to determine its crystallographic structure. We will see that lattice and magnetic properties are intertwined in a special way in \bcso\ and a profound and detailed knowledge of the crystal structure is crucial to unravel the underlying ordering phenomena.

Finger \textit{et al.} employed single-crystal X-ray diffraction and Raman spectroscopy to investigate the structure and vibrational properties of \bcso, which had been inadvertently synthesized during the search for Tl-based high-temperature superconductors. Indeed, Sheng and Hermann \cite{sheng1988} heated oxide mixtures in open silica containers. A reaction between the oxides and the silica produced a thin coating composed of two silicate phases: one appeared magenta under transmitted light, the other turquoise. The magenta phase, identified as \bcso, was found to be body-centered tetragonal, with lattice parameters \textit{a} = 7.042 \AA and \textit{c} = 11.133 \AA, and the space group determined as \textit{I$\bar{4}$m2} \cite{finger1989}.

Raman spectroscopy revealed a peak at 506 cm$^{-1}$, attributed to an in-plane breathing mode of the Si$_4$O$_{12}$ ring. Additionally, the presence of Cu$^{2+}$ dimers was identified. As we shall see later, this initial erroneous structure and space group assignment misled subsequent inelastic neutron scattering studies (INS), which incorrectly concluded that the inter-dimer magnetic exchange interaction is ferromagnetic \cite{sasago1997}.

A more detailed structural analysis was conducted by Sparta et al. \cite{sparta2004}. They observed superstructure reflections indicating that the room-temperature unit cell is four times larger than previously reported \cite{finger1989, mckeown1997}. This superstructure suggested a possible structural phase transition occurring above room temperature. The correct space group for the room-temperature phase of \bcso\ was identified as \textit{I4$_1$/acd}. Furthermore, a transition to a higher-symmetry phase with space group \textit{I4mmm} was confirmed at $T = 610$ K.

The discovery of split oxygen positions in the Cu$^{2+}$ environment led to a doubling of the unit cell and to a reassignment of the inter-dimer exchange interaction as antiferromagnetic. This revision established \bcso\ as a promising candidate for the realization of a magnon Bose-Einstein condensate \cite{jaime2004, sebastian2005, sebastian2006a}. The complete and accurate assignment of exchange interactions was finally achieved by Mazurenko \textit{et al.} \cite{mazurenko2014}, but we reserve the discussion for later.

\begin{figure}[ht]
\includegraphics[scale=.38]{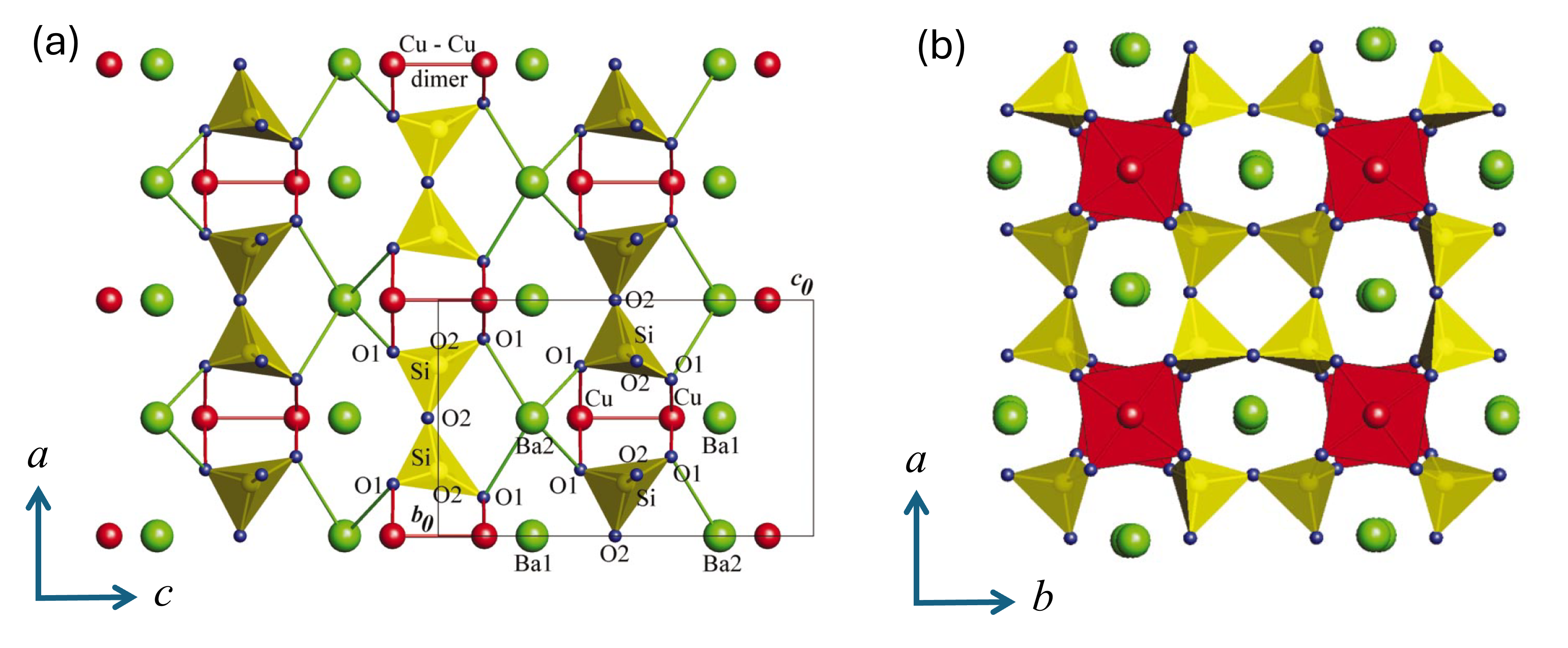}
\caption{(a) Room-temperature structure of \bcso\ in the originally/incorrectly assigned space group \textit{I$\bar{4}$m2}, according to Finger \textit{et al.} \cite{finger1989} in a projection onto the \textit{ac}-plane. Here, all Cu dimers are identical, and the unit cell indicated by the thin line is small.  (b) Projection of a Cu$_2$Si$_4$O$_{12}$ layer onto the \textit{ab}-plane. The green spheres represent the Ba atoms, the yellow tetrahedra the SiO$_4$ groups, and the red planar squares the CuO$_4$ groups. Model in the now widely accepted room temperature space group \textit{I4$_1$/acd}, with doubling of the unit cell. Reprinted with permission from Ref. \cite{sparta2004}. (Copyright {\copyright}2004 International Union of Crystallography)}
\label{fig:2}  
\end{figure}

The lattice properties of Han Purple continued to stimulate research for years. Building on this interest, Samulon \textit{et al.}, from the Stanford team, reported high-resolution X-ray diffraction experiments on single crystals over the temperature range of 16 to 300\,K. The data, shown in Fig.\,3(a), provide clear evidence of a transition from the room-temperature tetragonal phase to an incommensurately (IC) modulated orthorhombic structure below approximately 100\,K. Inside the orthorhombic phase, the modulation is characterized by a resolution-limited reduced wavevector \textbf{q}$_{IC}$ = (0, 0.129\,±\,0.001, 0) in reciprocal lattice units along with its second and third harmonics. The phase transition is first order accompanied by significant hysteresis.

At the time, it was difficult to determine precisely how the intradimer and interdimer exchange constants are affected by this structural modulation. Samulon et al. assumed that both interactions are modulated to some extent. They concluded that the spin Hamiltonian describing the system is more complex than originally presumed. As we will see later, follow-up INS experiments by R\"{u}egg \textit{et al.} \cite{ruegg2007}, performed with enhanced energy resolution, revealed multiple magnon modes—clear evidence of magnetically inequivalent dimer sites. These findings, along with NMR results by Kr\"{a}mer \textit{et al.} \cite{kramer2007, kramer2013}, pointed to a spatially modulated amplitude of the Bose-Einstein condensate.

\begin{figure}[ht]
\includegraphics[scale=.35]{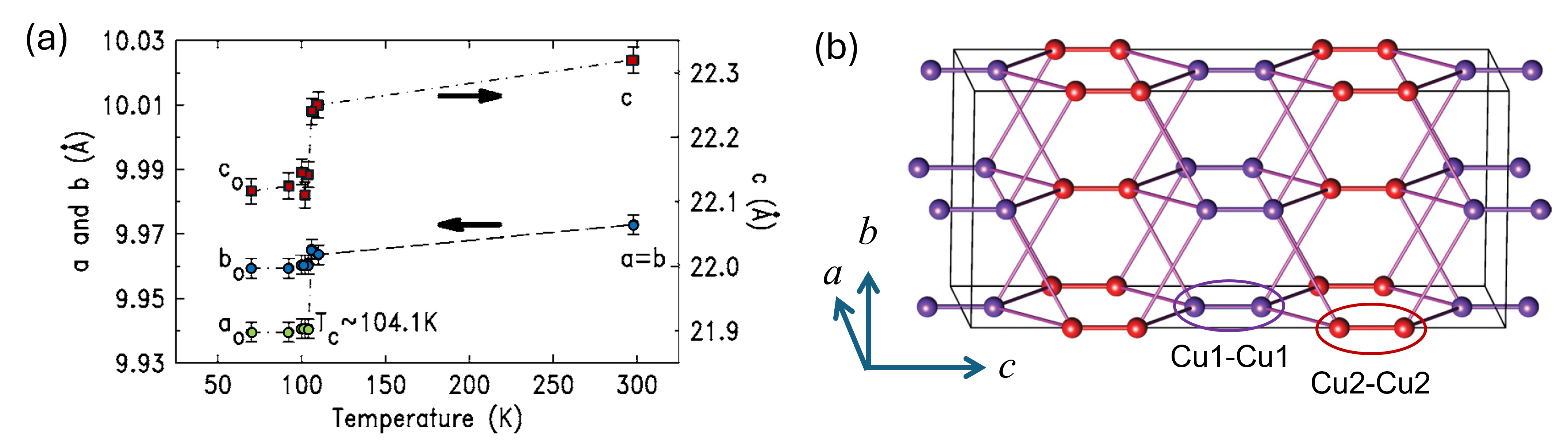}
\caption{(a) Lattice parameters of \bcso\ vs temperature on warming. Reproduced with permission from \cite{samulon2006} (Copyright {copyright}2006 American Physical Society). (b) Schematic representation of the crystal structure of the low-temperature phase with only the Cu-Cu dimers shown. For the interlayer coupling paths, Cu1-Cu2 (shown in thin lines in the right figure), there are always two pairs of slightly inequivalent bond distances. Reproduced with permission from  \cite{sheptyakov2012} (Copyright {\copyright}2012 American Physical Society).}
\label{fig:3}  
\end{figure}

Quantitative crystallographic evidence for lattice modulation along the $c$-axis in \bcso\ required significant additional effort. About six years later, the low-temperature crystal structure of \bcso\ was investigated using a combination of high-resolution synchrotron X-ray and neutron powder diffraction techniques. These studies revealed that the structure is, on average (ignoring the incommensurate modulation), orthorhombic, with the most probable space group \textit{Ibam}. Sheptyakov \textit{et al.} concluded that the Cu$-$Cu dimers form two types of layers with distinctly different interatomic distances, as shown schematically in Fig.\,3(b). Subtle structural changes also affect the partially frustrated interlayer Cu$-$Cu exchange paths. These results support the interpretation of low-temperature nuclear magnetic resonance (NMR) and INS data in terms of distinct dimer layers \cite{sheptyakov2012}.

Table\,1 summarizes the current understanding of the lattice parameters of the crystal structure of pure \bcso: it is tetragonal with space group \textit{I4mmm} at temperatures above 610\,K (high-temperature phase), tetragonal with space group \textit{I4$_1$/acd} for temperatures in the range $\sim$105\,K\,$<T<610$\,K (room-temperature phase), and orthorhombic with space group \textit{Ibam} below $T\,\sim\,90\,$K, where the lattice is modulated along the $c$-axis, resulting in at least two distinct Cu-dimer environments (see Fig.\,4). The structural phase transition in the 90$-$105$\,$K range is strongly hysteretic and likely first-order in nature. Remarkably, it can be fully suppressed by chemically substituting a few percent of Ba atoms with Sr \cite{vanwell2016, puphal2016}.

As we will see in a later section, this substitution—and the resulting suppression of the lattice modulation—played a crucial role in resolving a long-standing debate about the origin of the dimensional crossover from three-dimensional to two-dimensional (BEC) of magnons. The crossover occurs as temperature falls below 1\,K at the critical magnetic field $\mu_0H_c\,\simeq\,23\,$T, where the energy gap between the spin-singlet ground state and the lowest spin-triplet state closes.

Collegial interactions and healthy competition among diverse research teams across multiple institutions and facilities were instrumental in unraveling this phenomenally complex, yet deeply fascinating, challenge in materials science.

\begin{figure}[ht]
\includegraphics[scale=.35]{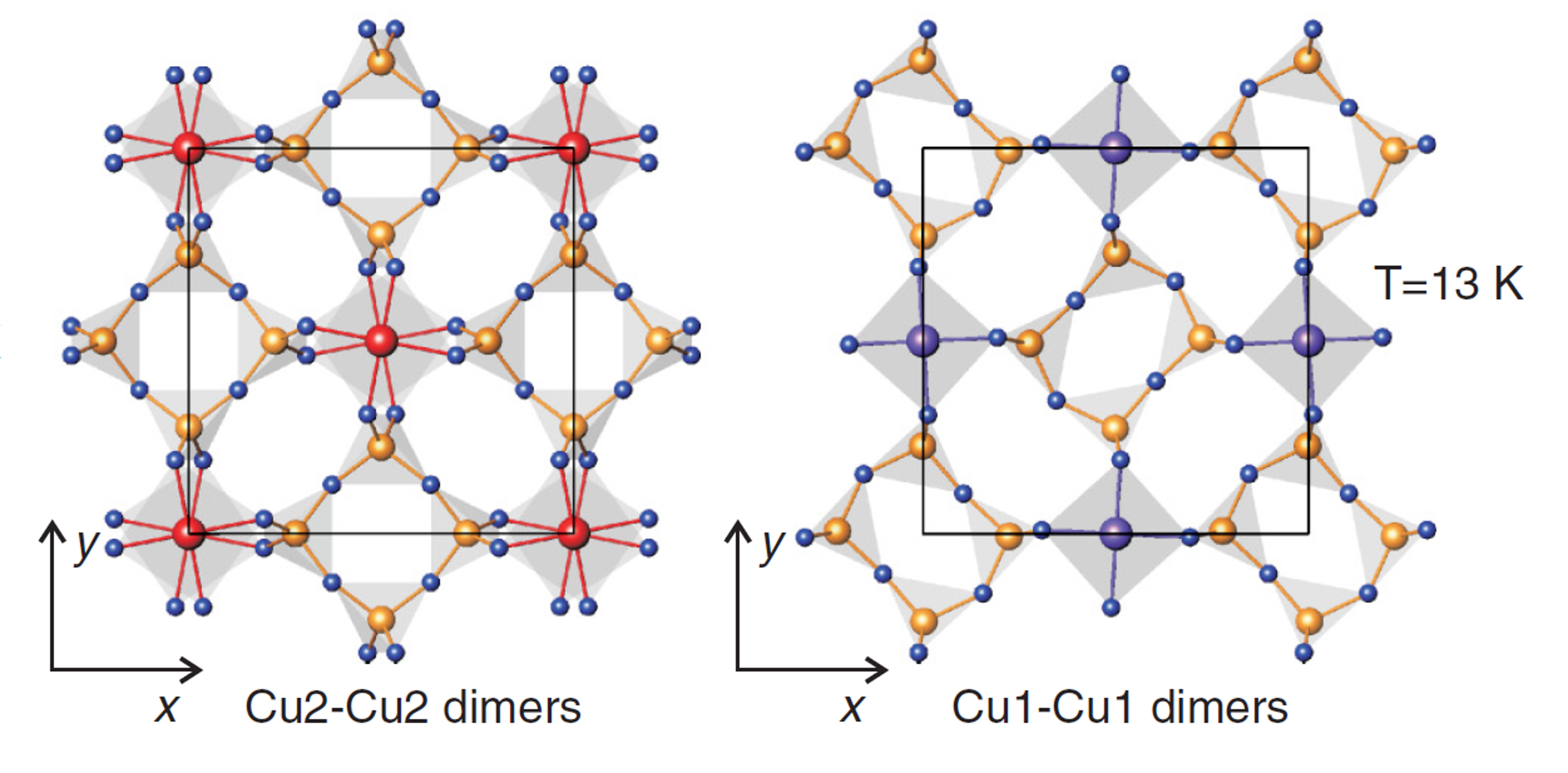}
\caption{Two types of Cu-Cu dimer layers at low temperature as refined from the neutron data taken at 13 K. The Cu2-Cu2 distance is 2.701(6) \AA ; the Cu1-Cu1 distance is 2.774(6) \AA. Red and blue spheres are copper, small blue spheres are oxygen, orange spheres are Si, Ba atoms, the spacer between copper bilayers, are not displayed. Reproduced with permission from  \cite{sheptyakov2012} (Copyright {\copyright}2012 American Physical Society).}
\label{fig:4}  
\end{figure}

\begin{table}[ht]
\caption{\bcso\ structure, space group,  and lattice parameters as discussed over the years}
\label{tab:2}       
%
%
\begin{tabular}{p{3cm}p{1cm}p{2.6cm}p{2cm}p{2.5cm}}
\hline\noalign{\smallskip}
reference & & space group & temperature & \textit{a,c} / \textit{a,b,c} (\AA)  \\
             \noalign{\smallskip}\svhline\noalign{\smallskip}

Finger \textit{et al.} (1989) &\cite{finger1989}&  Tetragonal \textit{I$\bar{4}$m2} & RT & 7.042, 11.133\\
Sparta \textit{et al.} (2004) &\cite{sparta2004}&  
 Tetragonal \textit{I4mmm} & $>\,$610\,K & 7.1104, 11.175\\
 &&  Tetragonal \textit{I4$_1$/acd} & 293\,-\,610\,K & 10.0091, 22.467\\
Samulon \textit{et al.} (2006) &\cite{samulon2006}& Tetragonal \textit{I4$_1$/acd} & RT & 9.97, 22.32\\
  & & Orthorombic & $T\lesssim$\,89\,-\,107\,K& 9.94, 9.96, 22.15\\
Sheptyakov \textit{et al.} (2012) & \cite{sheptyakov2012}& Tetragonal \textit{I4$_1$/acd} & RT & 9.97, 22.239\\
  && Orthorombic  \textit{Ibam} & $T$\,=\,13\,K & 9.951, 9.967, 22.239\\

\noalign{\smallskip}\hline\noalign{\smallskip}
\end{tabular}
\end{table}

\section{Magnetism and the formalism of BEC of spin degrees of freedom in quantum magnets}

\subsection{Spin-gap ground state controlled by magnetic fields and field-induced order}

Conventional AC magnetic susceptibility measurements on \bcso\ were first performed by Sasago \textit{et al.} \cite{sasago1997} on single crystals grown by the traveling molten floating zone method discussed in section 2.2. The measurements revealed an activated behavior of the susceptibility below 40\,K, indicative of a spin-singlet ground state with a spin gap. No evidence of a magnetic phase transition was observed within the measured temperature range. The raw susceptibility data (uncorrected for paramagnetic impurities) are shown as symbols in Fig.\,5(a). The experimental $\chi(T)$ data quantitatively agree with the theoretical prediction for an isolated dimer model, featuring a singlet-triplet gap of $\Delta = 4.1(0.03)$\,meV, as illustrated by the solid line in Fig.\,5(a).

The dimer ground state in \bcso\ was confirmed by INS experiments, performed on a 10\,x\,4\,x\,4\,mm$^3$ single crystal grown by the same floating-zone method \cite{jaime2004}. Inelastic constant-$Q$ scans at 3.5\,K revealed a gap excitation at $\hbar\omega \approx\,$4.5\,meV, consistently observed across the entire ($h,0,l$) plane [Fig.\,5(b)]. This gap energy was in reasonable agreement with the value inferred from susceptibility measurements using the isolated dimer model. 

Nevertheless, interdimer interactions induce a small but significant dispersion of the dimer excitations. Sasago \textit{et al.} \cite{sasago1997} reported no detectable dispersion along the $c$-direction, consistent with the quasi-two-dimensional, layered structure of \bcso. In contrast, along the $a$-direction, a finite bandwidth in the dimer modes was observed, shown as blue circles in Fig.\,5(c) for the 3.5\,K data.

Importantly, the dispersion bandwidth was found to be strongly temperature dependent. At 50\,K, the bandwidth along the (h,0,1.5) direction is significantly reduced compared to that at 3.5\,K. The experimental data were analyzed using a bilayer Heisenberg Hamiltonian incorporating three exchange constants: $J_1$ (intradimer), $J_2$ (interdimer within the plane), and $J'_2$ (interlayer coupling). A least-squares fit to the data at 3.5\,K yielded $J_1 = \Delta = 4.41(0.02)\,$meV, $J_2 = -0.19(0.03)\,$meV, and $|J_1 /J'_2|\simeq 24 \gg 1$, suggesting a ferromagnetic (FM) interdimer coupling.

\begin{figure}[ht]
\includegraphics[scale=.54]{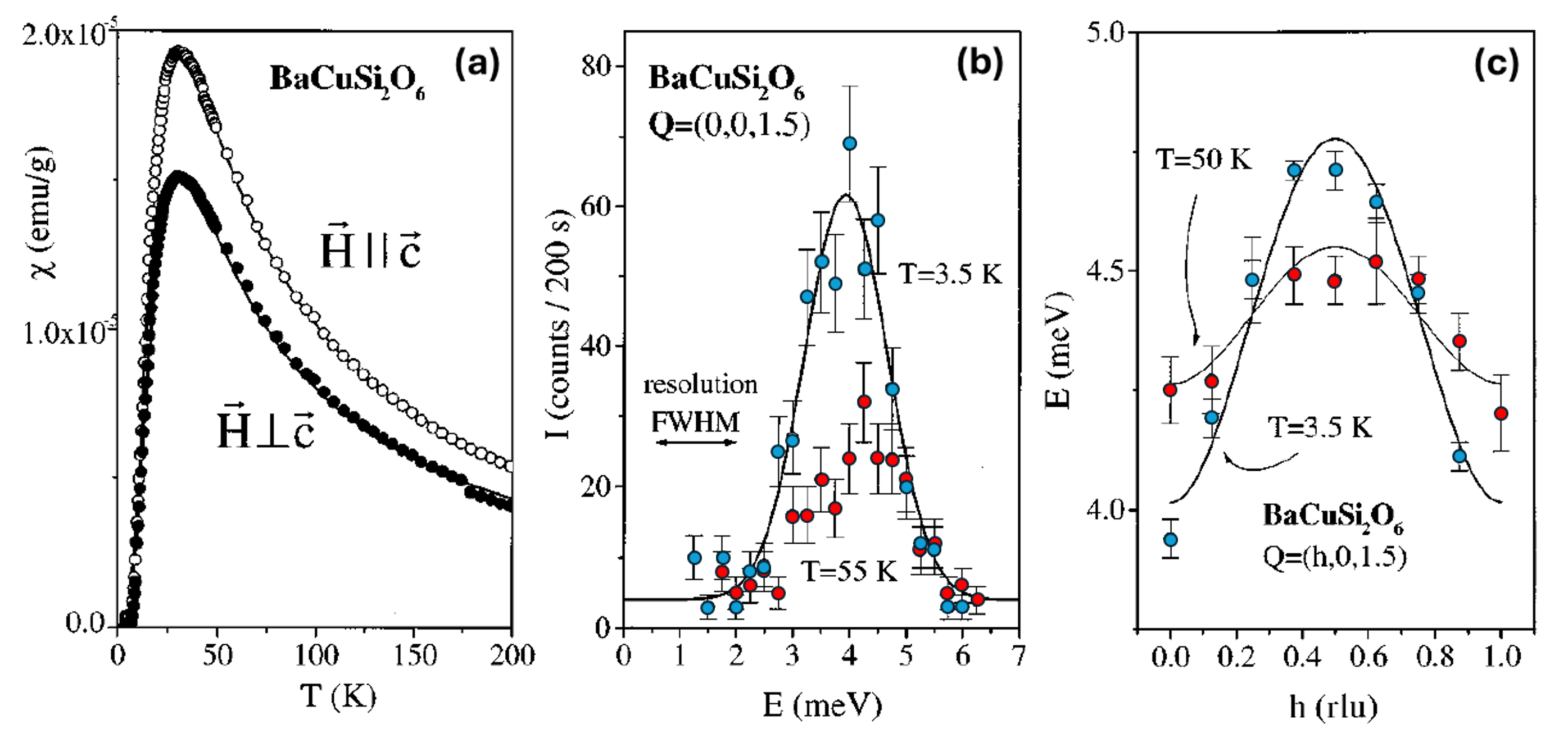}
\caption{(a) Magnetic susceptibility of a \bcso\ single crystal (symbols). The solid lines represent theoretical fits based on the isolated dimer model. (b) Representative constant-\textbf{Q} scans in \bcso, showing the magnetic gap excitation centered at $\hbar\omega = 4.5$\,meV. Solid lines are Gaussian fits to the data.
(c) Dispersion of the gap excitation measured along the ($h,0,1.5$) direction. The solid lines are fits to the experimental data, following the model described by Sasago \textit{et al.} \cite{sasago1997}. Reproduced with permission from \cite{sasago1997} (Copyright {\copyright}1997 American Physical Society)}
\label{fig:5}  
\end{figure}

However, this interpretation was soon challenged by magnetization measurements in pulsed magnetic fields up to 60\,T, high enough to close the 4.41\,meV spin gap. These high-field experiments provided alternative insights into the nature of interdimer interactions in \bcso.
Figure\,6(a) displays the low temperature magnetization vs magnetic field (red line) for a single crystal sample of \bcso\ when the magnetic field is applied along the crystallographic $c$-axis. A critical field $\mu_0H_{c1} \simeq 23\,$T marks the point where the magnetization becomes non-zero, reaching saturation at $\mu_0H_{c2} \simeq 49\,$T. The increase of magnetization between these is roughly linear within the scattering of the experimental data. Various physical properties measured in field sweeping mode as well as at constant magnetic field reveal a dome centered at 36\,T that extends to almost 4\,K \cite{jaime2004}. The mid point corresponds well to the intradimer exchange interaction, and the field span of 26\,T matches reasonably well the dispersion determined by Sasago et al. The linear behavior of magnetization between well-separated critical fields, however, is at odds with a FM interdimer exchange. Indeed, only a repulsion between field-induced excitations can explain the data, and this fact led to the revision of the modeling of the system.

A strong indication for a true phase transition is provided by specific heat measurements as a function of temperature for constant magnetic fields, as shown in detail in Fig.\,6(b). At zero magnetic field, the specific heat is featureless down to the lowest temperatures. As soon as a finite external field along the $c$-axis is applied, the specific heat increases due to the reduction in the singlet-triplet energy gap. A small anomaly then develops for $\mu_0H = 28\,$T at $T = 2.7\,$K that moves to higher temperatures as the magnetic field increases. This $\lambda$-shaped anomaly, which is a first hint for a second order phase transition, grows with magnetic field. Once the middle of the magnetization ramp is reached at $\mu_0H = 36\,$, the transition temperature and the size of the $\lambda$ anomaly start to decrease. No phonon contribution
was subtracted from the specific heat data.
The inset of Fig.\,6 displays the data at 37\,T after subtraction
of a small exponential contribution [activated
energy equal to 3.13\,meV] from the $S_z$ = 0 triplet
components that becomes apparent only above 8\,K. In
addition, a Debye phonon contribution with $\Theta = 350\,$K
was subtracted. The solid line is the result of a quantum
Monte Carlo calculation after a finite-size scaling
to the thermodynamic limit. 

\begin{figure}[ht]
\includegraphics[scale=.38]{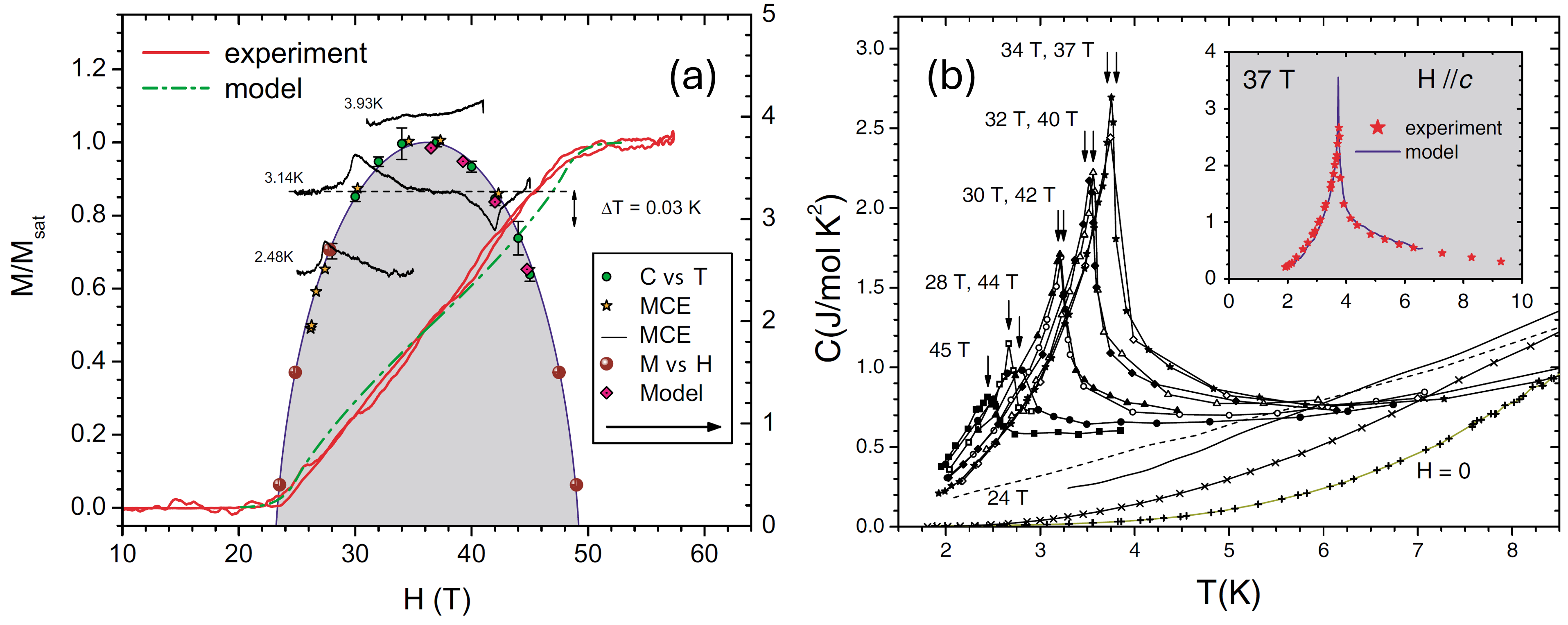}
\caption{(a) Left $y$-axis: magnetization normalized
to the saturation value ($M = M_{sat}$) vs magnetic field along the
$c$-axis, measured at 1.5\,K (red line). Results for the model
discussed in the text (green line). Right $y$-axis: transition
temperature from specific heat vs temperature, magnetocaloric
effect (MCE), and magnetization vs field data. Black lines are
the sample temperature measured while sweeping the magnetic
field quasi-adiabatically. (b) Specific heat vs temperature at constant
magnetic fields $H$. A low temperature $\lambda$ anomaly is evident at
$H \leq 28\,$T. The anomaly first moves to higher temperatures
with increasing magnetic fields, it reaches a maximum at $\mu_0H = 
36\,$T, and then decreases for $37<\mu_0H<45\,$T. Inset: specific
heat vs $T$ at $\mu_0H = 37\,$T after subtraction of a small contribution
due to the Sz = 0 triplet level only relevant at higher temperatures,
and phonons. Also displayed (inset) is the result of our Monte
Carlo calculation (solid black line). Reproduced with permission from \cite{jaime2004} (Copyright {\copyright}2004 American Physical Society).}

\label{fig:6}  
\end{figure}

\subsection{Implications of BEC formalism in the hard-core boson language}

The connection between magnetic spin systems and gases of itinerant particles capable of condensation became clear with the development of two key mathematical transformations introduced several decades ago. The first, proposed by Jordan and Wigner \cite{jordan1928}, applies to one-dimensional systems. It defines a nonlocal mapping between  $S=1/2$ spin operators and creation/annihilation operators for spinless fermions, a method extensively discussed in the literature \cite{zapf2014}. The Jordan-Wigner transformation is crucial for understanding the ground state properties and low-energy excitations of one-dimensional magnets. It also reveals that in one dimension, and for systems with short-range interactions, the distinction between fermionic and bosonic statistics can effectively be absorbed into the interaction terms.

About twenty years later, Matsubara and Matsuda introduced a simpler and more general mapping between  $S = 1/2$ spin operators and hard-core bosons \cite{matsubara1956}. Unlike the Jordan-Wigner transformation, the Matsubara-Matsuda mapping is local and applies in any spatial dimension because spin operators on different lattice sites naturally obey bosonic commutation relations. This mapping establishes a formal analogy between quantum magnetism and gases of hard-core bosons, where different types of magnetic ordering correspond to distinct condensed phases of bosons. In the hard-core boson language, the lowest energy singlet state (for the Cu$^{2+}$-Cu$^{2+}$ dimer in \bcso) maps into an empty boson state $\vert\uparrow\downarrow\rangle \rightarrow \vert0\rangle$, and the first exited state into occupied single boson state $\vert\uparrow\uparrow\rangle \rightarrow \vert1\rangle$.

In this framework, the applied magnetic field plays the role of the chemical potential, the longitudinal magnetization, $M_z$, corresponds to the boson density, and the square of the transverse magnetization, $M_{xy}^2$, serves as the order parameter \cite{jaime2004}. Moreover, axial (uniaxial) symmetry in the magnetic system translates into boson number conservation, enabling a direct analogy with Hubbard-type models. Continuing with the analogy, uniaxial symmetry manifests as a free precession of the $M_{xy}$ component, which in turn becomes the gapless BEC Goldstone mode. This perspective highlights the competition between kinetic energy of excitations (favoring boson itinerancy and BEC condensation) and potential energy or interactions (favoring boson localization that leads to spin-textured ground states).

Uniaxial symmetry ensures that the Hamiltonian commutes with $M_z$, preserving U(1) symmetry under spin rotations about the z-axis. However, real materials can have weak symmetry-breaking interactions due to spin-orbit coupling, dipole-dipole interactions, or Dzyaloshinskii-Moriya terms that pin the $M_{xy}$ component along an easy axis within the $xy$-plane. While these effects can relax metastable states by allowing $M_z$ to fluctuate, their influence is often negligible above tens of millikelvin, especially in crystals with high rotational symmetry and carefully chosen spin dimer structures.

Along the above lines, the ground state of the quantum paramagnet is boson-empty ($\vert0\rangle$) and features a gap for single boson creation. Increasing the applied magnetic field reduces this gap, and at the lower critical field $H_{c1}$ the gap closes, allowing bosons to populate the system and (if itinerancy is dominant) condense coherently into a BEC. This corresponds to the onset of $XY$-AFM magnetic order perpendicular to the applied field, where $\vert0\rangle$ and $\vert1\rangle$ coexist dynamically, forming a coherent extended state across the entire system.

As the field continues to increase, the boson density grows, but the hard-core repulsion (due to the two-level nature of the spins) limits occupancy to one boson per site. At the upper critical field $H_{c2}$, all spins become fully polarized along the field direction, corresponding to one boson $\vert1\rangle$ per site — a state akin to a Mott insulator in the bosonic picture. Thus, $H_{c1}$ marks the transition from a gap paramagnet to an ordered BEC-like phase, and $H_{c2}$ signals the transition from the ordered phase to a fully polarized state. The intrinsic particle-hole symmetry of this description manifests as a field-symmetric phase diagram as shown in Fig.\,7.

The QPTs (i.e., field-induced transitions at $T = 0$) at $H_{c1}$ and $H_{c2}$ belong to the BEC universality class characterized by a dynamical exponent $z = 2$, reflecting the quadratic dispersion of excitations near these points. Between $H_{c1}$ and $H_{c2}$, the system exhibits a dome-shaped region of $XY$ ordering, accompanied by a linear Goldstone mode dispersion due to spontaneous U(1) symmetry breaking. This ordered phase can be destabilized either by suppressing the amplitude of the order parameter (as in BEC transitions) or through enhanced phase fluctuations, leading to different critical behaviors such as the O(2) universality class with $z = 1$. 

Currently, a number of systems have been studied that reveal symmetry and magnetic properties consistent with the above phenomenology as discussed by Zapf \textit{et al.} \cite{zapf2014}, and displayed in Fig.\,7(left panel). Different temperature regimes in the temperature-magnetic field phase diagram are shown in Fig.\,7(right panel) ranging from the thermodynamic phase transition at the top of the dome, to the intermediate quantum-correlations dominated BEC regime characterized by universal behavior below $T = T_U \simeq T_c^{max}/4$ governed by dispersion relations and dimensionality (as discussed above) \cite{kawashima2004}, to the lowest temperature regime below $T = T_{bs}$ where the physical nature of the material deviates from the BEC formalism due to unavoidable broken symmetries (bs) in real crystals. In the case of \bcso, $T_{bs}$ is estimated to be in the few milliKelvin range \cite{sebastian2006b}, allowing for approximately a two-decades temperature range for the observation of universality between $T_{bs} \sim 10$\,mK and $T_c^{max}/4 \sim 1$\,K.  

\begin{figure}[ht]
\includegraphics[scale=.44]{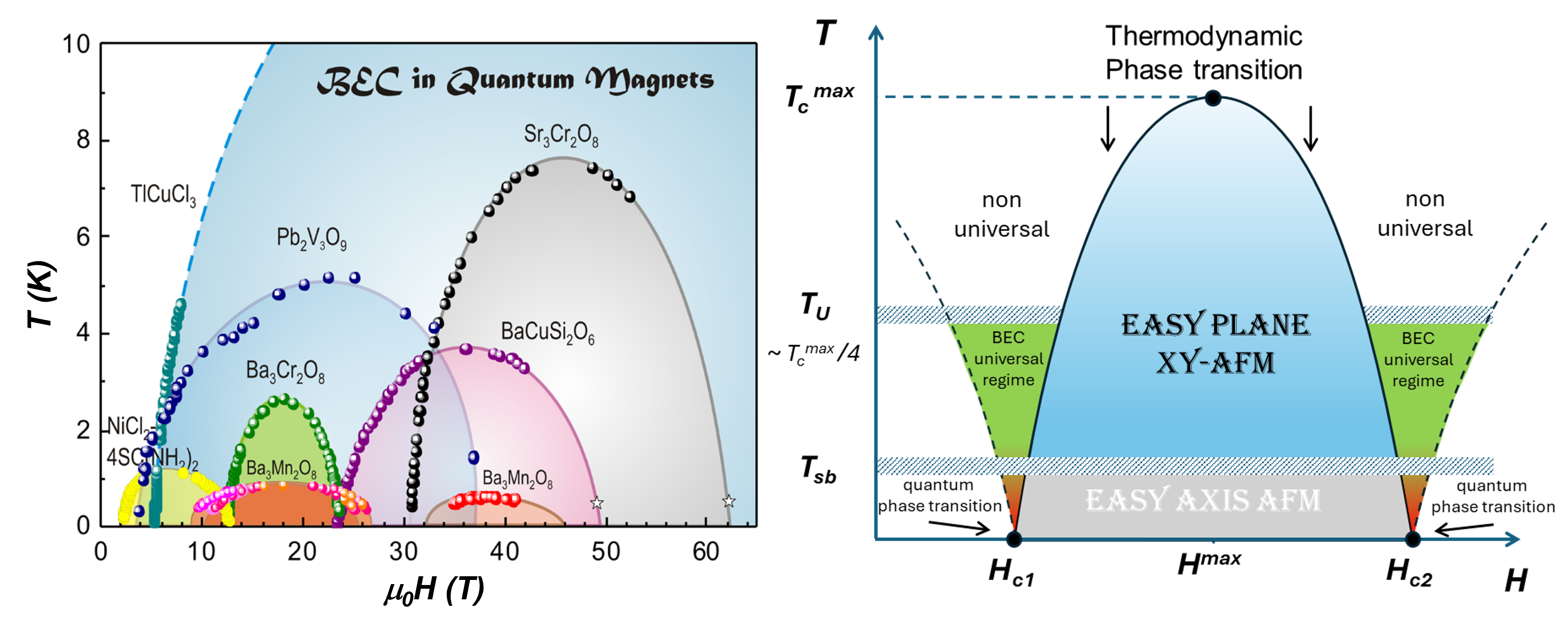}
\caption{(left panel) Experimental (T,H) phase diagram for several quantum magnet systems studied in the context of BEC. (right panel) Conceptual (T,H) phase diagram for a quantum magnet with a field-induced quantum critical point belonging to the BEC universality class. The region of magnetic ordering is delimited by the critical fields $H_{c1}$ and $H_{c2}$ and the maximum ordering temperature $T_c^{max}$ in the (T,H) plane. The energy scale $T_{sb}$ is the lower limit for observing a BEC QCP, and is given by uniaxial symmetry-breaking terms below which discrete symmetry, such as for instance easy-axis AFM, becomes important to the QCP and to the magnetic state inside the dome. $T_U$ is the maximum temperature expected for universal properties that are derived from low temperature approximations to the boson distribution function, such as $T^{3/2}$ scaling of thermodynamic properties for a $d$ = 3 BEC quantum critical point. Reprinted with permission from \cite{zapf2014} (Copyright {\copyright}2014 American Physical Society).}

\label{fig:7}  
\end{figure}

\subsection{Universal regime, critical exponents, and Goldstone mode}

The proximity in magnetic field to the field-induced quantum critical point (QCP) in a BEC system is expected to be related to the ordering temperature $T_c$ by a power law $T_c \propto (H - H_{c1})^{z/d}$, which can be expressed in reduced form 

\begin{equation}
t = f(h)(1 - h)^{z/d} 
 \label{Eq:1}     
\end{equation}

with parameters $t = T_c /T_c^{max}$, and $h = H^{max} - H / (H^{max} - H_{c1})$. $H^{max}$, and $T_c^{max}$ represent the point on the phase boundary halfway between $H_{c1}$ and $H_{c2}$, and $H_{c2}$ is the field at which the magnetization saturates. $z$ is the dynamical exponent, $d$ is the system dimensionality, and $f(h)_{h=1}$ is finite.\footnote{Eq. (1), a relatively simple equation of profound significance in the study of universal behavior, i.e., behavior independent of the microscopic details of the system, in solid state physics. It is likely the main responsible for fueling two decades of intense experimental and theoretical interest in \bcso , hence we include it here. } The mean-field critical exponent $\nu = 2/3$ ($z = 2$, $d = 3$) is characteristic of the three-dimensional (3D) BEC universality class, and describes the scaling behavior of a 3D dilute interacting Bose gas near the QCP. It must be noted that the dispersion relation for a BEC of magnons candidate has been experimentally studied \cite{ruegg2003, ruegg2004} in TlCuCl$_3$ by means of neutron scattering experiments under magnetic fields and applied pressures. It was confirmed that the dispersion relation is quadratic when the critical field $\mu_0H_{c1} \simeq 5.7$ T is approached, hence $z$ = 2. On the other hand, when the QCP is induced by applied pressure, the dispersion relation turns linear, and $z$ = 1 as expected. These crucial experiments provided early evidence for a Goldstone mode in TlCuCl$_3$, a gapless excitation considered a universal hallmark of continuous symmetry breaking. The quest in the case of \bcso\ is significantly more challenging due to the 4\,x larger magnetic fields required to reach the QCP at $\mu_0H_{c1} \simeq 23$\,T, which places this system outside the experimental window currently available for neutron scattering techniques.

The next best thing to do in the process of experimentally verifying the accuracy of the Bose-Einstein condensation of magnons approximation for \bcso\ is to study the universal regime via the temperature dependence of the critical magnetic field $H_c(T)$, via physical properties that can be measured in large DC magnetic fields as well as pulsed magnetic fields. Good examples of these are the specific heat $C_p(T,H$ = const$)$, magnetocaloric effect $T(H)$, and longitudinal magnetization $M(H,T$ = const$)$ \cite{jaime2004,jaime2010}, displayed in Fig.\,6(a,b), torque magnetometry \cite{sebastian2005, harrison2006, sebastian2006a}, and NMR experiments \cite{horvatic2005}, and electron spin resonance ESR \cite{zvyagin2006}. Indeed, in addition to confirm the ordered state, as seen by anomalies in the physical properties in Fig.\,6, the temperature dependence can be computed with a minimalist Hamiltonian and quantum Monte Carlo techniques \cite{jaime2004} and, last but not least, the low temperature limit of phase boundaries contrasted against Eq.\,1 describing the universal regime, i.e. regime independent of microspopic details of the system.

However, it is extremely difficult to verify Eq.\,1 experimentally. This is due to three very important issues, namely that i) usually just a few data points are available at the lowest end of the temperature scale
due to equipment limitations, and that ii) the critical exponent varies with temperature as the experimental probe transits the crossover from non-universal to universal regimes below $T_U$, compare Fig.\,7(left panel), resulting in a polynomial fit of the curve ($\mu_0H_c$, $T_c$). Using Eq.\,1 yields different exponent values for different temperature windows utilized for the fit. And most importantly, iii) $H_{c1}(T=0)$ that is critical for the fit, is utterly unknown to the experimentalist. More often than not, simple polynomial fits are severely limited by the above issues and do not produce reliable values for the critical exponent $\nu$.

\begin{figure}[ht]
\includegraphics[scale=.70]{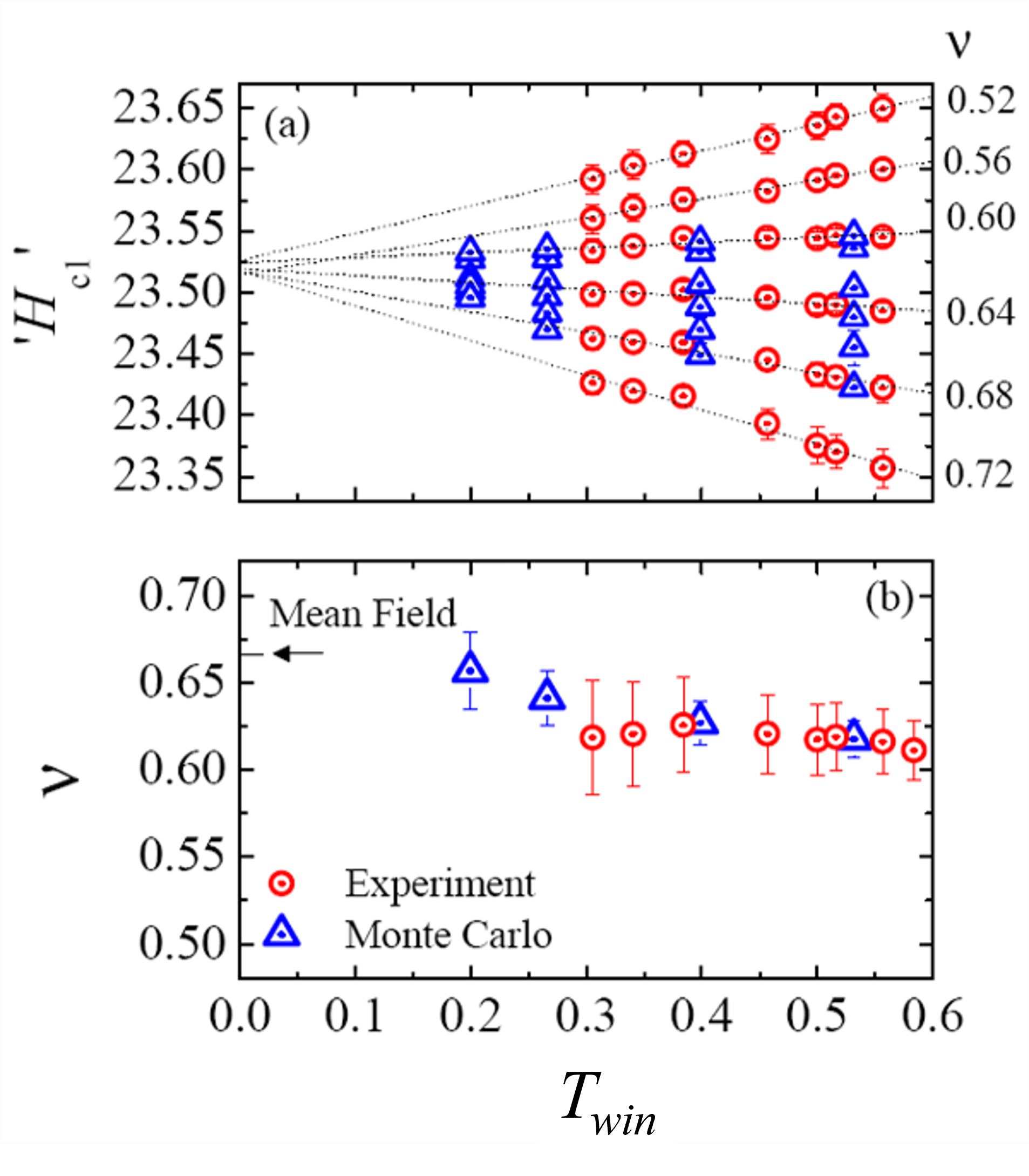}
\caption{(a) Circles represent estimates of \uno~ obtained from
fitting the lowest few experimental points on the ($\mu_0H_c$,$T_c$) phase boundary of \bcso\ in a window of increasing size $T_{win}$, to Eq. 1 for different fixed values of $\nu$. The $x$ axis $T_{win}$ labels the highest reduced temperature of the fit window. The dotted lines show the linear convergence of \uno~ values at $T_{win}$ = 0. Triangles represent estimates of \uno~ similarly obtained from Monte Carlo simulation data for corresponding fixed values of $\nu$, and similar convergence is observed. (b) Circles represent estimates of $\nu$ from fitting the lowest few experimental
points on the phase boundary in a window of increasing
size $T_{win}$, to Eq. 1 with \uno~ = 23.52 T determined from (a). The error bars are due to experimental uncertainty in determining the values of
$\mu_0H_c$. Triangles represent estimates of $\nu$ from a similar fit to Monte
Carlo simulation data. The dotted line is a guide to the eye, illustrating the approach of the Monte Carlo simulation to the mean field
value as $T_{win} \rightarrow$ 0. Reproduced with permission from \cite{sebastian2005} (Copyright {\copyright}2005 American Physical Society).}

\label{fig:8}  
\end{figure}

Sebastian \textit{et al.} used single crystal samples of \bcso\ grown by a flux-growth technique to measure the specific heat $C_p$, magnetocaloric effect (MCE), and torque magnetometry in a DC magnet furnished with a $^3$He cryostat to determine the ($\mu_0H_c$, $T_c$) curve with high precision down to 0.61\,K, almost one order of magnitude lower than $T_{max}$ \cite{sebastian2005}. Then, they used an analysis technique with an independent determination of $H_{c1}$ from the experimental data. This fixed value was used to perform a one-parameter fit to $\nu$, which therefore required a smaller number of data points in a limited range to obtain a statistically significant fit. This technique, referred to as the sliding window method, displayed and explained in Fig.\,8 yielded the lowest experimentally accessible temperature window containing nine data points down to 0.61 K to obtain $\mu_0H = 23.52$\,T and a value of $\nu = 0.63 \pm 0.03$ in good agreement with theoretical expectations \cite{nohadani2004, nohadani2005, wang2006, giamarchi2008}.

Shortly after, these experiments and analysis were extended by the same team more than a decade lower in temperature using a dilution refrigerator. The results are displayed in Fig.\,9. Here, an unexpected crossover to a different value of the critical field $\mu_0$\uno = 23.17 T and critical exponent $\nu$ = 1 are uncovered, below $T \sim$ 0.5\,K. Sebastian \textit{et al.} speculated that the exponent crossover is caused by a dimensional crossover from 3D to 2D as the temperature is lowered. Such dimensional reduction, while anti-intuitive at first when considering that quantum fluctuations are expected to diverge spatially as the temperature approaches the absolute zero, was argued to be rooted in geometrical magnetic frustration between stacked Cu$^{2+}$ bilayers in the crystal structure. A theoretical analysis found this interpretation physically sound, endorsing the idea of a first observation of a geometrical frustration-induced dimensional reduction \cite{batista2007, schmalian2008, rosch2007a}. Indeed, they demonstrate that the dimensionality of the BEC-QCP in \bcso\ can be $d = 2$ when the interlayer coupling is frustrated. However, this coupling is also relevant for changing the thermodynamic phase transition from  Berezinskii-Kosterlitz-Thouless (BKT) at low temperatures to the 3D-$XY$ universality class at high temperatures. Intriguingly, these results explain quantitatively and without free parameters, the dimensional reduction manifested in the measured quantum critical exponents of \bcso. However, further theory considerations using a generalized bond-operator method by R\"{o}sch \textit{et al.} \cite{rosch2007b} calculate the low temperature magnetic properties in the paramagnetic and antiferromagnetic phases. Based on the available experimental data on \bcso, they propose that a scenario with inequivalent layers and imperfect frustration is realized in this material, likely with an additional modulation of the interlayer coupling along the $c$-axis.    

\begin{figure}[ht]
\includegraphics[scale=.50]{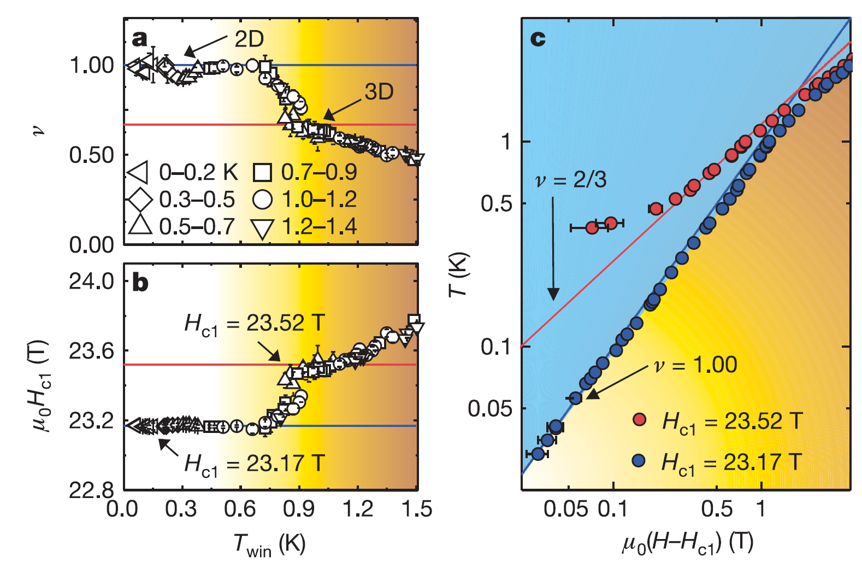}
\caption{Crossover from 3D to 2D BEC critical exponent. \textbf{a}, Values of the critical exponent $\nu$ obtained from fitting experimental points on the phase boundary in a sliding window centered at Twin (K). A two-parameter least squares regression of Eq.1  with $\nu$ and critical field $H_{c1}$ varying is used to fit the data. Error bars show the standard errors in the least squares fit. The window size varies from 0.05 to 1.4 K, as indicated by different symbols. The data approach $\nu$ = 2/3 in the intermediate regime, and there is a
distinct crossover toward $\nu$ = 1 before the QCP is reached. \textbf{b}, Estimates of $\mu_0H_{c1}$ obtained along with $\nu$ during the fit. $\mu_0H_{c1}$ approaches a transient value of 23.52 T in the intermediate regime, and crosses over to the true low
temperature value of 23.17 T before the QCP is reached. The shading reflects the crossover toward the $\nu$ = 1 exponent as the QCP is approached both as a function of field and temperature. The dark yellow shading indicates the high temperature (field) non-universal regime, the bright yellow the intermediate regime, and the light yellow the 2D regime. \textbf{c}, Best fits to the phase boundary in the intermediate and low temperature regimes represented on a logarithmic scale. The solid lines show that the data in the intermediate temperature range are consistent with the values of $\mu_0H_{c1}$ = 23.52 T and $\nu$ = 2/3, which does not fit the lower temperature points; whereas data in the lower temperature range are consistent with $\mu_0H_{c1}$ = 23.17 T and $\nu$ = 1, which does not fit the higher temperature points. A crossover is observed from one regime to the other in the temperature range 0.65K $< T_{win} <$ 0.9 K. Reproduced from Sebastian \textit{et al.} \cite{sebastian2006a} (Copyright {\copyright}2006 Springer Nature).}

\label{fig:9}  
\end{figure}

\subsection{Structural phase transition and lattice modulation}

Following the experimental studies on critical exponents, and inspired by theoretical work described in the previous section, a rather comprehensive set of experiments was conducted to better understand the low-temperature structural properties of \bcso. Among them, Samulon \textit{et al.} \cite{samulon2006}  obtained results of high-resolution X-ray diffraction experiments on single crystal samples in the temperature range 16 to 300\,K, see also discussion in section\,3). The data they collected show unambiguous evidence of a crystallographic transition from the room-temperature tetragonal phase into an incommensurately modulated orthorhombic structure below $\sim$100\,K. The authors first suggest in this work that the lattice modulation could impact the symmetry of the magnetic ground state.

\begin{figure}[ht]
\includegraphics[scale=.60]{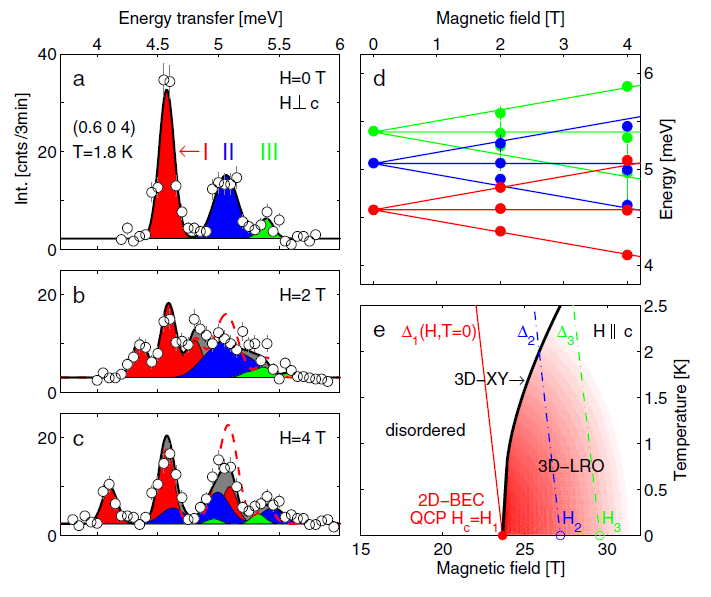}
\caption{(a)-(c) INS spectra at three magnetic field
values in \bcso\ for \textbf{Q} = (0.6, 0, 4), $T = 1.8$\,K, and $k_f$ = 1.3 \AA$^{-1}$. (d) Zeeman splitting of triplet modes following the procedure described in the text. (e) Schematic phase diagram
around the QPT, intensity of red shading indicates the qualitative degree of $c$-axis modulation of the BEC order parameter. Reproduced with permission from R\"{u}egg \textit{et al.} \cite{ruegg2007} (Copyright {\copyright}2007 American Physical Society).}
\label{fig:10}  
\end{figure}

On a different front, R\"{u}egg \textit{et al.} investigated the spin dynamics by INS experiments on \bcso\ single crystals with high energy resolution and observed multiple magnon modes, indicating clearly the presence of magnetically inequivalent dimer sites.
The more complex spin Hamiltonian revealed in this study leads to a distinct form of magnon Bose-Einstein condensate phase with a spatially modulated condensate amplitude, as seen in Fig.\,10 \cite{ruegg2007}. Simultaneously, Kr\"{a}mer \textit{et al.} \cite{kramer2007, stern2014} carried out a comprehensive $^{63,65}$Cu and $^{29}$Si NMR study in the magnetic field range 13$-$,26\,$T$ and at temperatures as low as 50\,mK. NMR data in the spin gap phase reveal that below 90\,K, alongside the lattice parameter modulation, different intradimer exchange couplings and different gaps ($\Delta B/\Delta A = 1.16$) exist in every second plane along the $c$-axis, in addition to a planar  IC modulation. $^{29}$Si spectra in the field-induced magnetic ordered phase reveal in this study that close to the quantum critical point at \uno\, =\,23.35\,T, the average boson density of the Bose-Einstein condensate is strongly modulated along the $c$-axis with a density ratio for every second plane equal to 5. An IC modulation of the local density is also present in each plane. (See Fig.\,11). Furthermore, Sheptyakov \textit{et al.} \cite{sheptyakov2012} investigated the low-temperature crystal structure with high-resolution synchrotron X-ray and neutron powder diffraction techniques.  The lattice has been found to be, on average (ignoring the IC modulation), orthorhombic, with the most probable space group \textit{Ibam}. They confirmed that Cu-Cu dimers in this material are forming two types of layers with distinctly different interatomic distances. Subtle changes also modify the partially frustrated interlayer Cu-Cu exchange paths. The present results corroborate the interpretation of low-temperature NMR and INS data in terms of distinct dimer layers.

\begin{figure}[ht]
\includegraphics[scale=.40]{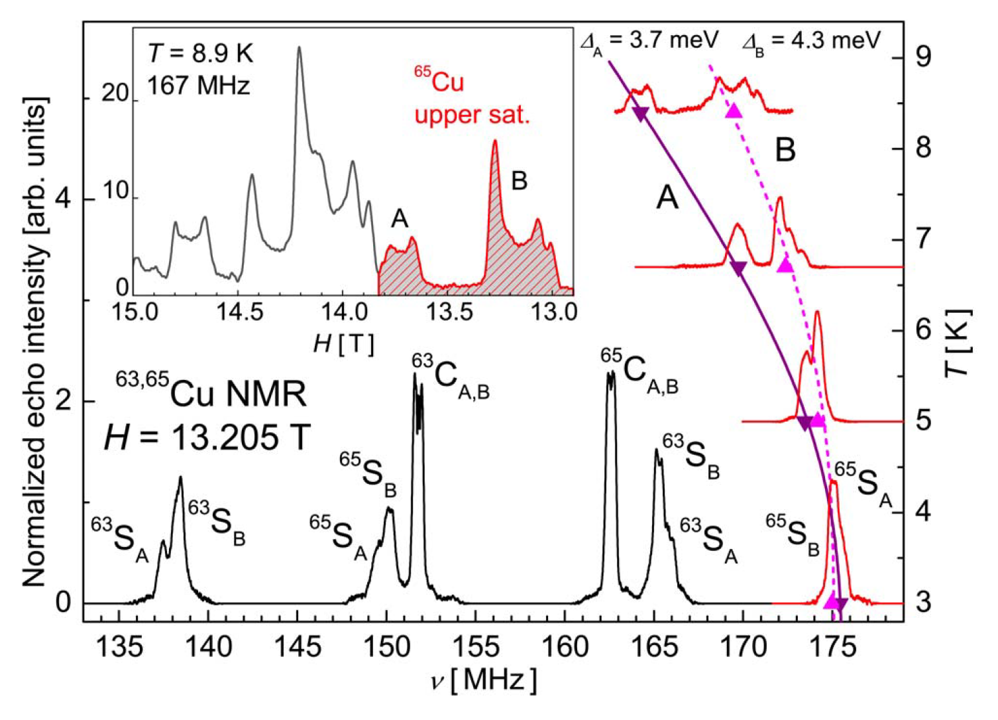}
\caption{$^{63,65}$Cu NMR spectra of \bcso\ in
the spin gap phase, well below the critical field. The T dependence of
the high-frequency satellite line clearly reveals two different copper
sites. From their shifts, the two corresponding gap values have
been determined. Inset: field sweep spectrum that reveals the IC
nature of the line shape for each of the two sites. Shading separates
the contribution of the $^{65}$Cu high-frequency satellite from the rest of the spectrum. The analysis of the latter part confirms that the observed line shape has a pure magnetic origin. Reproduced with permission from \cite{kramer2007} (Copyright {\copyright}2007 American Physical Society).}
\label{fig:11}  
\end{figure}

Intense theoretical work followed suit these findings. Laflorencie and Mila \cite{laflorencie2009}, complementing the work of Batista \textit{et al.} on one hand, and R\"{o}sch and Vojta\cite{rosch2007a} on the other, investigated the field-induced exotic criticality observed in the frustrated compound \bcso\ using a frustrated model with two types of bilayers inspired by the NMR work. With a semiclassical treatment of the effective hard-core boson model, they show that perfect interlayer frustration leads to a 2D-like critical exponent $\nu = 1$ without logarithmic corrections and to a 3D low temperature phase with different but non-vanishing triplet populations in both types of bilayers. In this way, they show that all these experimental data can in fact be reconciled in the context of a model including the two types of bilayers and perfect frustration. As mentioned already before, R\"{o}sch and Vojta were inclined towards an inequivalent bilayers interpretation of the 2D regime later in the discussion \cite{rosch2007b}).

\begin{figure}[ht]
\includegraphics[scale=.34]{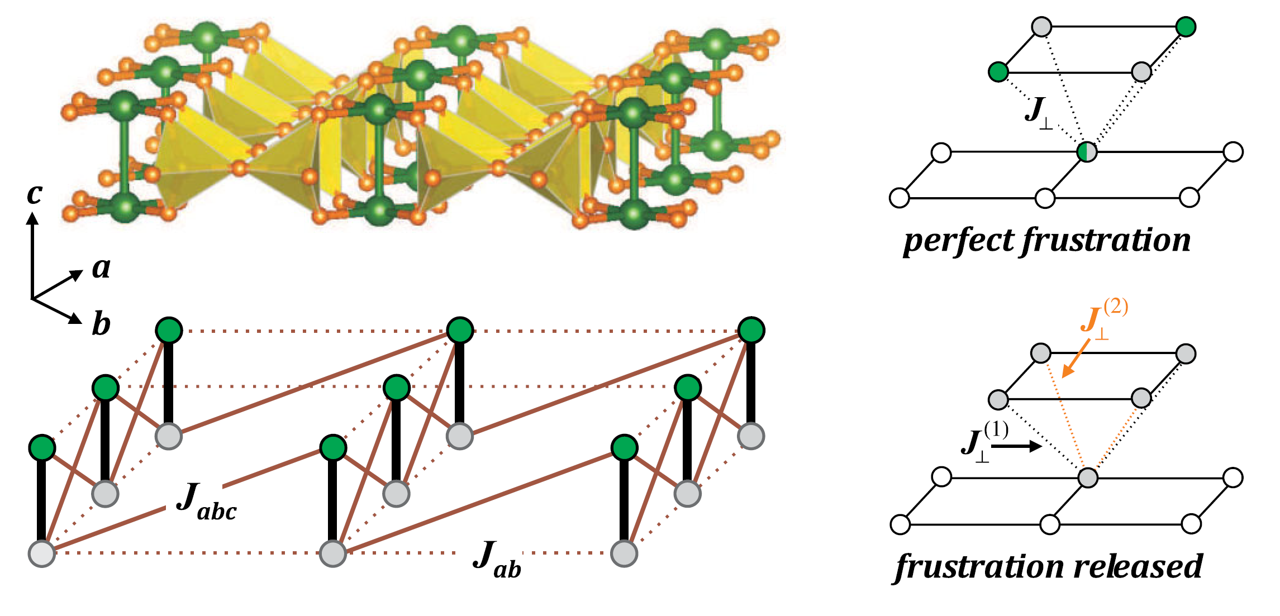}
\caption{(Left panel) crystal structure of the magnetic bilayer and the relevant magnetic model with the FM in-plane order driven by the AFM interdimer coupling $J_{abc}\gg J_{ab}$. Green (dark) and gray (light) circles denote different spin directions. Right panel: different regimes of the interlayer order depending on the in-plane magnetic order. The AFM in-plane order leads to a perfect frustration (top). The FM in-plane order lifts the frustration (bottom). $J^{(1)}_{\perp}$ and $J^{(2)}_{\perp}$ are two different interlayer couplings in the low-temperature structure of \bcso. Both are weak and FM. Crystallographic plots are done using the VESTA software. Reproduced with permission from \cite{mazurenko2014} (Copyright {\copyright}2014 American Physical Society).}

\label{fig:12}  
\end{figure}

\section{Strong interplay between lattice and magnetism}

Using extensive density-functional calculations Mazurenko \textit{et al} \cite{mazurenko2014} reassessed the magnetic couplings in \bcso\ and found no evidence for inter-bilayer frustration. The resulting magnetic model comprises two types of nonequivalent spin dimers, in excellent agreement with the $^{63,65}$Cu NMR data as shown in Fig.\,11, They further argue that leading interdimer couplings connect the upper site of one dimer to the bottom site of the contiguous dimer, and not the upper-to-upper and bottom-to-bottom sites, as assumed previously. This finding is verified by INS data and implies the complete lack of frustration between the layers of spin dimers in \bcso, thus challenging existing theories of the origin of two-dimensional-like Bose-Einstein condensation of magnons in this compound. 

S. Allenspach \textit{et al.} \cite{allenspach2020} show with high-resolution neutron spectroscopy experiments that the effective intrabilayer interactions are ferromagnetic, thereby excluding frustration. They further explain the apparent dimensional reduction by establishing the presence of three magnetically inequivalent bilayers, with ratios 3:2:1, whose differing interaction parameters create an extra field-temperature scaling regime near the QCP with a nontrivial but nonuniversal exponent. They demonstrate by detailed quantum Monte Carlo simulations with the deduced magnetic interaction parameters  that all the measured properties of \bcso\ can be accounted for, opening the way to a quantitative understanding of nonuniversal scaling in any modulated layered system. It must be said that the intense collegial scientific work carried out in this compound across theory and experimental teams to fully understand the nature of the magnetic ground state is by all means extraordinary, and led to enormous progress. 

\begin{figure}[ht]
\includegraphics[scale=.35]{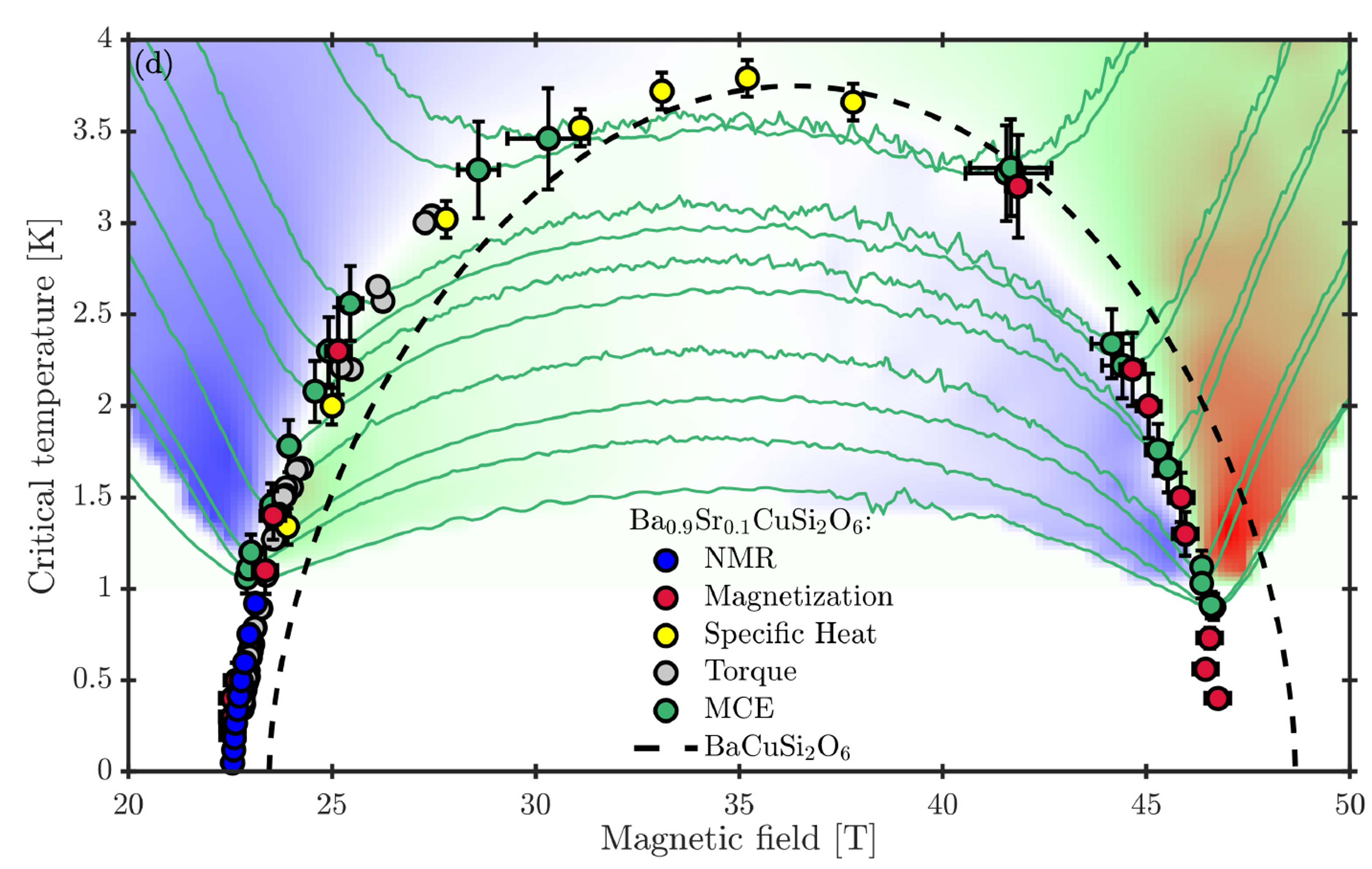}
\caption{Field-temperature phase diagram of Ba$_{0.9}$Sr$_{0.1}$CuSi$_2$O$_6$ obtained by collecting experimental data from magnetization, specific heat, magnetic torque, magnetocaloric effect, and NMR. Solid lines are lines of constant entropy obtained from our MCE measurements. Color contours represent the magnetic Gr\"{u}neisen parameter,  $\mu_0H = -(\partial M/\partial T )/C_p = -(\partial T/\partial H)S/T$, which shows a sharp change of sign at the phase boundary. The black dashed line shows for comparison the phase boundary of \bcso. Reproduced with permission from  \cite{allenspach2021} (CCA 4.0 International).}
\label{fig:13}  
\end{figure}

Two developments followed the discussion that both focused on controlling the magnetic structure  dimensionality by means of lattice deformation, a last remaining barrier to the full understanding of BEC of magnons in \bcso\ when just one bilayer type exists.

The first one took place in the group led by H. Tanaka at the Tokyo Institute of Technology, pursuing a modified version of \bcso\ with dicoupled Cu bilayers that shift the BEC properties towards the 2D limit. They synthesized single crystals of composition Ba$_2$CuSi$_2$O$_6$Cl$_2$ and investigated their quantum magnetic properties \cite{okada2016} both experimentally and theoretically. In fact, the crystal structure of this compound is found to be closely related to \bcso\ with a singlet ground state, and an excitation gap of $\Delta$/$k_B$ = 20.8\,K. The magnetization curves for two different field directions almost perfectly coincide when normalized by the $g$ factor, indicating that the magnetic anisotropy is very small. However, the material was found to be fully magnetically frustrated in elastic neutron scattering experiments by Kurita \textit{et al.} \cite{kurita2019}. A more detailed discussion is found in Chapter 11. Magnetic excitations of the effective spin $S = 1/2$ dimerized magnet Ba$_2$CuSi$_2$O$_6$Cl$_2$ have been probed directly via INS experiments at temperatures down to 4\,K. They observed five types of excitation at 4.8, 5.8, 6.6, 11.4, and 14.0\,meV, which are all dispersionless within the resolution limits. The scattering intensities of the three low-lying excitations were found to exhibit different $Q$-dependencies. Detailed analysis, discussed in Chapter 12, has demonstrated that Ba$_2$CuSi$_2$O$_6$Cl$_2$ is a two-dimensional spin dimer system described only by a single bilayer. However, triplet excitations are localized owing to the almost perfect frustration of the interdimer exchange interactions. We conclude that making BEC behavior in \bcso\ more 2D by adding an intercalate seems difficult, since it changes the ground state more profoundly.

\begin{figure}[ht]
\includegraphics[scale=.40]{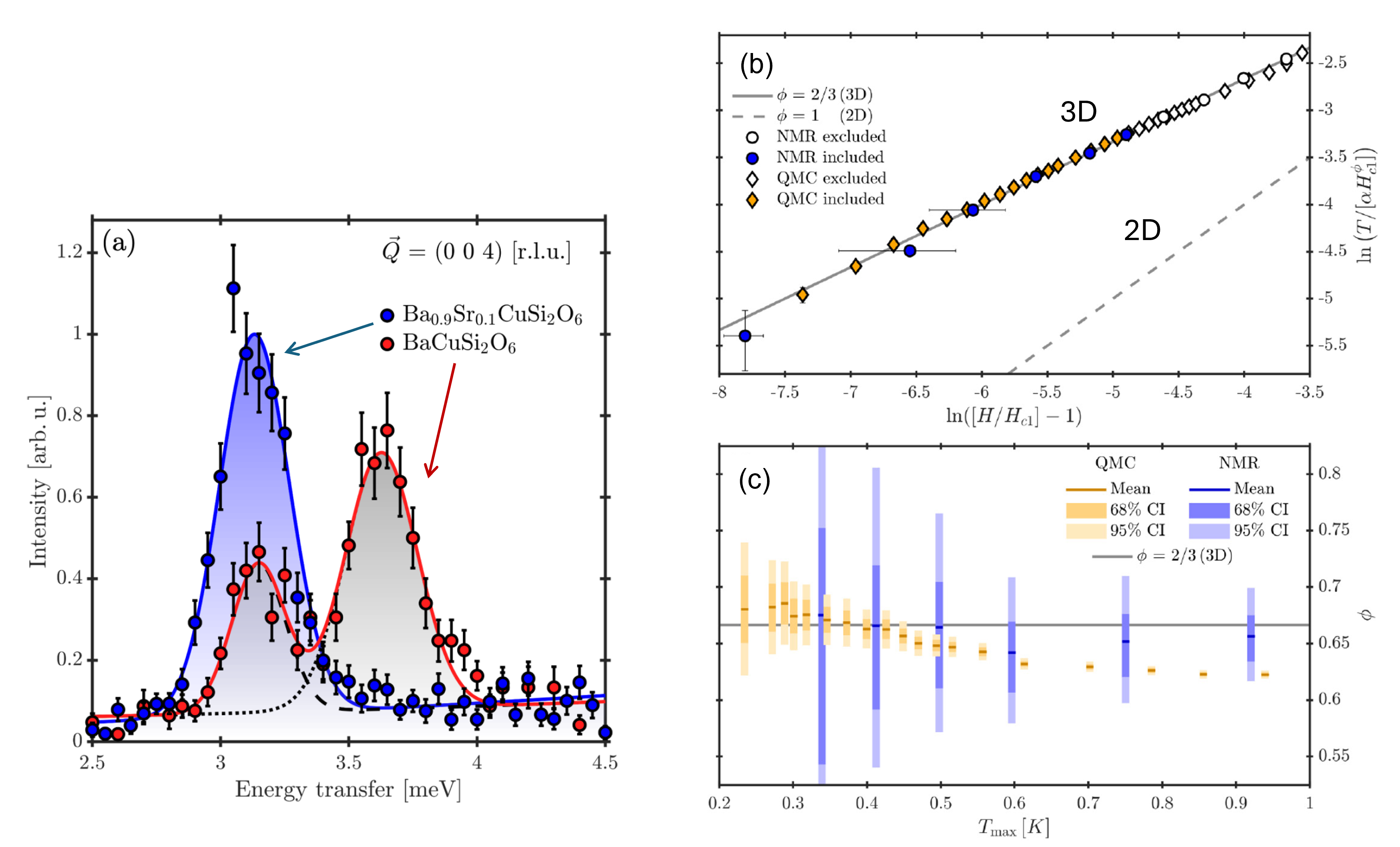}
\caption{(a) Scattered intensity as a function of energy transfer, measured on TASP at \textbf{Q} = (0 0 4) r.l.u. for Ba$_{0.9}$Sr$_{0.1}$CuSi$_2$O$_6$ at 1.5\,K (blue symbols) and \bcso\ at 1.6\,K (red). Shaded regions represent Gaussian fits used to extract
the triplon energy, linewidth, and intensity. (b) Phase-boundary points of Fig.\,13 shown on logarithmic axes. The gradients of both sets of points correspond to the critical exponent, $\nu$. For
comparison we indicate the cases $\nu = 1$ (2D, dashed) and $\nu = 2/3$ (3D, solid). (c) Evolution of $\nu$ as a function of the maximal temperature, $T_{max}$, up to which phase-boundary data points are included. Horizontal lines show the mean and shaded boxes the 68$\%$  and 95$\%$ confidence intervals (CI) of the marginal distribution of $\nu$ deduced by including only points below $T_{max}$; the 95$\%$ CIs for $T_{max}$ = 264\,mK lie at 0.465 and 0.923. The NMR symbols were obtained by including 5, 6, ..., 10 of the data points shown in Fig.\,13 and panel (b), the quantum Monte Carlo (QMC) symbols by including between 8 and 30 data points. Reproduced with permission from \cite{allenspach2021} (CCA 4.0 International).}
\label{fig:14}  
\end{figure}

Another route was taken by the group of Krellner at Goethe-Universit\"{a}t Frankfurt, inhibiting the lattice structural transition with slight doping and suppressing the $c$-axis modulation in \bcso.  This approach, also discussed in Chapter 3, succeeded with 10\,\% Sr doping \cite{vanwell2016, puphal2016, puphal2019} retaining the magnetic properties of the pure compound. Following work performed by collaborators at 18 institutions across 8 countries \cite{allenspach2020} on Ba$_{0.9}$Sr$_{0.1}$CuSi$_2$O$_6$ applied the full toolbox of experimental methods developed for \bcso\ to reveal the true BEC nature without lattice involvement. Zero-field magnetic excitations were measured by neutron spectroscopy, and the spin Hamiltonian was deduced. The magnetic field-induced transition was probed by magnetization, specific heat, torque, and magnetocaloric effect measurements in high pulsed and DC fields. Furthermore, NMR studies near $H_{c1}$ were carried out to the lowest temperatures. Fig.\,13 summarizes the results. With a Bayesian statistical analysis and large-scale Quantum Monte Carlo simulations, they demonstrated unambiguously that observable 3D quantum critical scaling with $\nu = 0.67^{+0.05}_{-0.07}$  is restored down to 200\,mK. Details of the analysis are presented in Fig.\,14. Although the impurity concentration is increased in Sr-doped \bcso, it is remarkable that no signatures of broken symmetry or magnetic easy axis ordering are observed in the investigated temperature range. We would like to mention that the critical fields \uno, and \dos\ are only slightly shifted to lower values as seen in Fig.\,13.

The results clearly establish \bcso\ as a quantum spin system in which the crystal lattice dictates the ground-state behavior, acting as the driving force behind the emergence of 2D magnetic excitations at the lowest temperatures, rather than as a merely passive witness to quantum magnetism$-$driven phenomena \cite{jaime2022}.

\section{Conclusions and outlook}

\bcso\ has proven to be a fascinating material to work on—from its early synthesis in ancient human cultures, to its serendipitous rediscovery during the search for high-temperature superconductors, to its pivotal role in directed research on Bose-Einstein condensation (BEC) in low-dimensional systems. Initially appearing as a textbook example of a U(1) magnetic symmetry system, and benefiting from its compatibility with high-quality single-crystal growth in materials science laboratories, \bcso\ inspired early experimental and theoretical efforts employing state-of-the-art tools.

In this chapter, we focus our discussion on the synthesis routes developed to date, the crystal structure, and magnetism. We presented the first unambiguous indications of a spin excitation gap small enough to be studied away from the interference of room-temperature phonons, and tuned with laboratory-accessible magnetic fields. Indeed, by closing this spin gap using applied magnetic fields, we revealed a thermodynamic phase transition in observables such as specific heat vs. temperature, a well-established proof of a thermodynamically stable state of matter. The archetypal lambda-like anomaly —hallmark of a second-order phase transition— was successfully reproduced using quantum Monte Carlo techniques and a minimalistic Hamiltonian containing only two exchange interactions: intra- and inter-dimer antiferromagnetic Cu$-$Cu coupling.

Due to the experimental difficulties to detect a Goldstone mode, which is considered key evidence for BEC at the field-induced quantum critical point at $\mu_0H_{c1} = 23\,$T via inelastic neutron scattering experiments, significant effort was invested in confirming the BEC-related universal behavior of physical properties near and along the phase boundary $H_{c1}(T)$. These studies illuminated the crucial roles of dimensionality, geometrical frustration (or lack thereof), and lattice symmetry in stabilizing the magnetic ground state, while also revealing the limitations of the BEC formalism in fully capturing the experimental phenomenology. Contrary to initial expectations, \bcso\ is not a "simple" square-coordination bilayer spin system. Instead, its degree of freedom of lattice reduces the free energy of the system through structural modulations—including along the $c$-axis—giving rise to not one but multiple coexisting spin excitation condensates. These condensates enrich the experimental response across multiple probes with several energy scales acting at once, making the interpretation and confirmation of BEC physics much more challenging.

Interestingly, when only one of these condensates is probed in isolation, the resulting physics is effectively two-dimensional—a rare and captivating consequence of strong magnetic and lattice correlations. Several aspects remain under active debate. On the one hand, the question remains unanswered if geometric frustration alone is sufficient to decouple the AFM-coupled bilayers and to generate a 2D BEC. On the other hand, it is still unresolved if there is any real spin compound system exhibiting strongly 2D BEC behavior in a single condensate.

Attempts to isolate the 2D limit via chemical substitution have thus far been unsuccessful, though it is a promising path for future exploration. In particular, a successful strategy to prevent the decoupling of Cu bilayers—and thus inhibit condensate segregation—enabled the realization of 3D universal behavior. This achievement required coordinated experimental and theoretical efforts across international teams. For now, we can conclude that this particular debate is settled: slightly Sr-doped \bcso\ represents a demonstrated realization of a 3D Bose-Einstein condensate of magnons, whereas the undoped compound continues to stand as a strong candidate for the rare manifestation of the 2D limit.  

We have admittedly left out many important details and contributions regarding the experimental and theoretical efforts to understand the ground-state magnetic properties of \bcso. Instead, we have focused on what can arguably be described as the backbone of the extensive body of work developed over the past two decades. Along this journey, new and improved techniques were developed to access the elusive nature of \bcso—an iconic member of the quantum magnet family, where the rare XY easy-plane magnetic state (i.e., BEC of magnons) is, albeit briefly, both experimentally realized and mathematically approximated.

Undoubtedly, these efforts have also helped refine the tools used to study and understand a wide range of other magnetic systems, including quantum spin liquid candidates, skyrmionic systems, magnetic supersolids, and altermagnets. These advances have recently galvanized the interest of the community in discovering the microscopic mechanisms that may one day unlock  applications.

\begin{acknowledgement}

 M. Jaime acknowledges support by the Cluster of Excellence QuantumFrontiers, funded by the Deutsche Forschungsgemeinschaft (DFG, German Research Foundation) under Germany's Excellence Strategy $-$ EXC 2123, Project-ID 390837967.

The authors are grateful for many years of insightful discussions and collaborations with colleagues in the magnetism community. We especially thank R. Stern for bringing \bcso\ to our attention as a "plan B" candidate for pulsed-field experiments at the National High Magnetic Field Laboratory, and E.W. Fitzhugh and H. Berke for revealing to us the fascinating archeological and anthropological roots of \bcso. We also acknowledge, among others, N. Harrison, C.D. Batista, N. Kawashima, S.E. Sebastian, V. Zapf, V. Correa, Ch. R\"{u}egg, S. Zvyagin, P. Sengupta, T. Kimura, G.S. Boebinger, S. Kr\"{a}mer, the late C. Berthier, J. Schmalian, B. Normand, F. Mila, M. Vojta, A. Tsirlin, H. Tanaka, T. Giamarchi, and P. Puphal for countless illuminating exchanges that helped shape this work. We are also grateful for the thoughtful skepticism of Art Ramirez, who still owes some of us a bottle of good wine!

\end{acknowledgement}


\begin{thebibliography}{99.}%
%
%
%


\bibitem{ma2006}Ma, Q., Portmann, A., Wild, F., and Berke, H.,: Raman and SEM Studies of man-made barium copper silicate pigments in ancient Chinese artifacts. Studies in Conservation, \textbf{51}, 81$-$98 (2006)
 
\bibitem{berke2006}Berke, H., : The invention of blue and purple pigments in ancient times. Chem. Soc. Rev. \textbf{36}, 15 (2007) 

\bibitem{fitzhugh1992}FitzHugh, E.W., Zycherman, L.A.: A Purple Barium Copper Silicate Pigment from Early China. Stud. Conserv. \textbf{37}, 145--154 (1992) doi.org/10.2307/1506342

\bibitem{qin2016}Qin, Y.,  Wang, Y.-H., Chen, X., Li, H.-M. Li, X.-L.: A Discussion on the Emergence and Development of Ancient Chinese Artificial Barium Copper Silicate Pigments from Simulation Experiments. Archaeometry \textbf{58}, 796$-$806 (2016) 

\bibitem{cormar2019} Corona-Mart\'{i}nez, D.A., J.Rend\'{o}n-Angeles, C., Gonzalez, L.A.,  Matamoros-Veloza, Z., Yanagisawa, K., Tamayo, A., Alonso, J.R.: Controllable synthesis of BaCuSi$_2$O$_6$ fine particles via a one-pot hydrothermal reaction with enhanced violet colour hue. Advanced Powder Technology \textbf{30}, 1473--1483 (2019)

\bibitem{berke2002}Berke, H.: Chemistry in Ancient Times: The Development of Blue and Purple Pigments. Angew. Chem. Int. Ed. \textbf{41}, 2483--2487 (2002)

\bibitem{wikipedia}Terracotta Army. Wikipedia, \url{https://en.wikipedia.org/wiki/Terracotta_Army}

\bibitem{rieck2015}Rieck, B., Pristacz, H., Giester, G.: Colinowensite, BaCuSi$_2$O$_6$, a new mineral from the Kalahari Manganese Field, South Africa and new data on wesselsite, SrCuSi$_4$O$_{10}$. Mineralogical Magazine. \textbf{79}, 1769--1778 (2015) doi:10.1180/minmag.2015.079.7.04

\bibitem{finger1989}Finger, L.W., Hazen, R.M., Hemley, R.J.: BaCuSi$_2$O$_6$: A  new cyclosilicate with four-membered tetrahedral rings. American Mineralogist \textbf{74}, 952--955 (1989)

\bibitem{sasago1997}Sasago, Y., Uchinokura, K., Zheludev, A., Shirane, G.: Four-membered silicate rings: Vibrational analysis of BaCuSi$_2$O$_6$ and implications for glass structure. Phys. Rev. \textbf{B56}, 3114--3121 (1997)

\bibitem{sparta2004}Sparta, K.M., Roth, G.: Reinvestigation of the structure of BaCuSi$_2$O$_6$ -- evidence for a phase transition at high temperature. Acta Cryst.  \textbf{B60}, 491--495 (2004) 

\bibitem{bouherour2001}Bouherour, S., Berke, H., Wiedemann, H-G.: Ancient Man-made Copper Silicate Pigments Studied by Raman Microscopy. Chimia \textbf{55}, 942--951 (2001) 

\bibitem{garcia2015}Garc\'{i}a-Fern\'{a}ndez, P., Moreno, M. Aramburu, J.A.,: Origin of the Exotic Blue Color of Copper-Containing Historical Pigments. Inorg. Chem. , \textbf{54}, 192$-$199 (2015)

\bibitem{jaime2004}Jaime, M., Correa, V.F., Harrison, N., Batista, C.D., Kawashima, N., Kazuma, Y., Jorge, G.A., Stern, R., Heinmaa, I., Zvyagin, S.A., Sasago, Y., Uchinokura, K.: Magnetic-Field-Induced Condensation of Triplons in Han Purple Pigment BaCuSi$_2$O$_6$. Phys.Rev.Lett. \textbf{93}, 087203 (2004) 

\bibitem{sebastian2006a}Sebastian, S.E., Harrison, N., Batista, C.D., Balicas L., Jaime, M.,  Sharma, P.A., Kawashima, N., Fisher, I.R.: Dimensional reduction at a quantum critical point. Nature \textbf{441} 617--620 (2006)

\bibitem{allenspach2022}Allenspach, S., Madsen, A., Biffin, A., Bartkowiak, M., Prokhnenko, O., Gazizulina, A., Liu, X., Wahle, R., Gerischer, S., Kempfer, S., Heller, P., Smeibidl, P., Mira, A., Laflorencie, N., Mila, F., Normand, B., R\"{u}egg, Ch.,: Investigating field-induced magnetic order in Han purple by neutron scattering up to 25.9 T. Phys. Rev. \textbf{B106}, 104418 (2022)

\bibitem{sebastian_thesis}Sebastian, S.E.,: Bose-Einstein condensation in spin dimer compounds, Ph.D. Thesis, Stanford University, (2006)

\bibitem{horvatic2005}Horvati\'{c}, M., Berthier, C., Tedoldi, F., Comment, A., Sofin M., Jansen, M., Stern, R.: High-Field NMR Insights into Quantum Spin Systems -- evidence for a phase transition at high temperature. Prog. Theor. Phys. Suppl.  \textbf{159}, 106--112 (2005) 

\bibitem{zvyagin2006}Zvyagin, S.A., Wosnitza, J., Krzystek, J., Stern, R., Jaime, M.,  Sasago, Y.,  Uchinokura K.: Spin-triplet excitons in the S = 1/2 gapped antiferromagnetic BaCuSi$_2$O$_6$: Electron paramagnetic resonance studies. Phys. Rev. \textbf{B73}, 094446 (2006) 

\bibitem{sebastian2005}Sebastian, S.E., Sharma, P.E., Jaime, M., Harrison, H., Correa, V., Balicas, L., Kawashima, N., Batista, C.D., Fisher, I.R.: Characteristic Bose-Einstein condensation scaling close to a quantum critical point in BaCuSi$_2$O$_6$. Phys. Rev. \textbf{B72}, 100404(R) (2005) 

\bibitem{samulon2006}Samulon, E.C., Islam, Z., Sebastian, S.E., Brooks, P.B., McCourt, Jr., M.K., Ilvsky, J., Fisher, I.R.: Low-temperature structural phase transition and incommensurate lattice modulation in the spin-gap compound BaCuSi$_2$O$_6$. Phys. Rev. \textbf{B73}, 100407(R) (2006) 

\bibitem{sebastian2006b}Sebastian, S.E., Tanedo, P., Goddard, P.A.,  Lee, S.-C., Wilson, A., Kim, S.,  Cox, S., McDonald, R.D., Hill, S., Harrison, N., Batista, C.D., Fisher I.R.: Role of anisotropy in the spin-dimer compound BaCuSi$_2$O$_6$. Phys. Rev. \textbf{B74}, 180401(R) (2006) 

\bibitem{jaime2010}Jaime, M.: Netsu Sokutei \textbf{37}, 26 (2010)

\bibitem{harrison2006}Harrison, N., Sebastian, S.E., Batista, C.D., Jaime, M., Balicas L., Sharma, P.A., Kawashima, N., Fisher, I.R.: Bose-Einstein condensation in BaCuSi2O6. Jou. Phys.: Conf. Series  \textbf{51}, 9--14 (2006)

\bibitem{sebastian2007}Sebastian, S.E., Harrison, N., Batista, C.D., Balicas L., Jaime, M.,  Sharma, P.A., Kawashima, N., Fisher, I.R.: BEC phase boundary in BaCuSi$_2$O$_6$. J. Magn. Magn. Mater. 310 (2007) e460$-$e462 Jou. Phys.: Conf. Series  \textbf{310}, e460--e462 (2007)


\bibitem{puphal2019}Puphal, P., Allenspach, S., R\"{u}egg, Ch., Pomjakushina, E.: Floating Zone Growth of Sr Substituted Han Purple:
Ba0.9Sr0.1CuSi2O6 Ba$_{0.9}$Sr$_{0.1}$CuSi$_2$O$_6$. Crystals \textbf{9}, 273 (2019)

\bibitem{chen2014}Chen, Y., Zhang, Y., Feng, S.: Hydrothermal synthesis and properties of pigments Chinese purple BaCuSi$_2$O$_6$ and dark blue BaCu$_2$Si$_2$O$_7$. Dyes and Pigments \textbf{105}, 167--173 (2014)

\bibitem{vanwell2016}van Well, N., Puphal, P., Whinger, B., Schefer, J., R\"{u}egg, Ch., Ritter, F., Krellner, C., Assmus, W.: Crystal Growth with Oxygen Partial Pressure of the BaCuSi$_2$O$_6$ and Ba$_{1-x}$Sr$_x$CuSi$_2$O$_6$ Spin Dimer Compounds. Crystal Growth and Design \textbf{16}, 3416--3424 (2016)

\bibitem{sheng1988}Sheng, Z.Z, Hermann, A.M.: Superconductivity in the rare-earth-free Tl-Ba-Cu-O system above liquid nitrogen temperature. Nature, \textbf{332},55-58 (1988). \textit{Idem}:  Bulk superconductivity at 120 K in the Tl-Ca/Ba-Cu-O system. Nature, \textbf{332}, 138-13 (1988)

\bibitem{mckeown1997}McKeown, D.A., Bell., M.I.: Temperature-dependent spin gap and singlet ground state in BaCuSi$_2$O$_6$. Phys. Rev. \textbf{B55}, 8357--8360 (1997)

\bibitem{mazurenko2014}Mazurenko, V.V., Valentyuk, M.V., Stern, R., Tsirlin, A.A.: Nonfrustrated Interlayer Order and its Relevance to the Bose-Einstein Condensation of Magnons in BaCuSi$_2$O$_6$. Phys.Rev.Lett. \textbf{112}, 107202 (2014) 

\bibitem{ruegg2007}R\"{u}egg, Ch., McMorrow, D.F., Normand, B., R{\o}nnow, H.M., Sebastian, S.E., Fisher, I.R., Batista, C.D., Gvasaliya, S.N.,  Niedermayer, Ch., Stahn, J.: Multiple Magnon Modes and Consequences for the Bose-Einstein Condensed Phase in BaCuSi$_2$O$_6$. Phys. Rev. Lett. \textbf{98}, 017202 (2007)

\bibitem{kramer2007} Kr\"{a}mer, S., Stern, R., Horvati\'{c}, M., Berthier, C.,  Kimura, T., Fisher I.R.: Nuclear magnetic resonance evidence for a strong modulation of the Bose-Einstein condensate in BaCuSi$_2$O$_6$. Phys. Rev. \textbf{B76}, 100406(R) (2007) 

\bibitem{kramer2013}Kr\"{a}mer, S., Laflorencie, N., Stern, R, Horvati\'{c}, M., Berthier, C., Nakamura, H., Kimura, T., Mila, F.: Spatially resolved magnetization in the Bose-Einstein condensed state of BaCuSi$_2$O$_6$: Evidence for imperfect frustration. Phys. Rev. \textbf{B87}, 180405(R) (2013)


\bibitem{sheptyakov2012}Sheptyakov, D.V., Pomjakushin, V.Y., Stern, R., Heinmaa, I., Nakamura, H., Kimura, T.,: Two types of adjacent dimer layers in the low-temperature phase of BaCuSi$_2$O$_6$. Phys. Rev. \textbf{B86}, 014433 (2012) 

\bibitem{jordan1928}Jordan, P., and E. Wigner, Z. Phys. \textbf{A 47}, 631 (1928)

\bibitem{zapf2014}Zapf, V., Jaime, M., Batista, C.D.: Bose-Einstein condensation in quantum magnets. Rev. Mod. Phys. \textbf{86}, 563--614 (2014)

\bibitem{matsubara1956}Matsubara, T., and H. Matsuda, Prog. Theor. Phys. \textbf{16}, 569 (1956)

\bibitem{kawashima2004}Kawashima, N.: Quantum Critical Point of the XY Model and Condensation of Field-induced Quasiparticles in Dimer Compounds. J. Phys. Soc. Jpn. \textbf{73}, 3219 (2004)

\bibitem{ruegg2003}R\"{u}egg, C., N. Cavadini, A. Furrer, H.-U. G\"{u}del, K. Kr\"{a}mer, H. Mutka, A. Wildes, K. Habicht, and P. Vorderwisch, Nature
\textbf{423}, 62 (2003).

\bibitem{ruegg2004}R\"{u}egg, C., A. Furrer, D. Sheptyakov, T. Str\"{a}ssle, K.W. Kr\"{a}mer, H.-U. G\"{u}del, and L. M\'{e}l\'{e}si, Phys. Rev. Lett. \textbf{93}, 257201 (2004)

\bibitem{nohadani2004}Nohadani, O., Wessel, S., Normand, B., Haas, S.,: Universal scaling at field-induced magnetic phase transitions. Phys. Rev. \textbf{B69}, 220402(R) (2004)

\bibitem{nohadani2005}Nohadani, O., Wessel, S., Haas, S.,: Quantum phase transitions in coupled dimer compounds. Phys. Rev. \textbf{B 72}, 024440 (2005)

\bibitem{wang2006}Wang, H.-T., Xu, B, Wang, Y.: Field-induced condensation of magnons and long range order in BaCuSi$_2$O$_6$. J. Phys.:Condens. Matter \textbf{18} 4719--4730 (2006)

\bibitem{giamarchi2008}Giamarchi, T., R\"{u}egg, Ch., Tchernyshyov, O.: Bose$-$Einstein condensation in magnetic insulators.: Nat. Phys. \textbf{4}, 198 (2008) 

\bibitem{batista2007}Batista, C.D., Schmalian, J., Kawashima, N., Sengupta, P., Sebastian N., Harrison, N., Jaime, M., Fisher I.R.: Geometric Frustration and Dimensional Reduction at a Quantum Critical Point. Phys. Rev. Lett. \textbf{98}, 257201 (2007) 

\bibitem{schmalian2008} Schmalian, J., Batista, C.D.,: Emergent symmetry and dimensional reduction at a quantum critical point. Phys. Rev. \textbf{B77}, 094406 (2008) 

\bibitem{rosch2007a} R\"{o}sch O., Vojta, M.,: Quantum phase transitions and dimensional reduction in antiferromagnets with interlayer frustration. Phys. Rev. \textbf{B76}, 180401(R) (2007) 

\bibitem{rosch2007b} R\"{o}sch O., Vojta, M.,: Reduced dimensionality in layered quantum dimer magnets: Frustration vs. inhomogeneous condensates. Phys. Rev. \textbf{B76}, 224408 (2007) 

\bibitem{stern2014}Stern, R., Heinmaa, I., Joon, E., Tsirlin, A.A.,  Nakamura, H., Kimura, T.: Low-Temperature High-Resolution Solid-State
(cryoMAS) NMR of Han Purple BaCuSi$_2$O$_6$. Appl. Magn. Res. \textbf{45}, 1253--1260 (2014)

\bibitem{laflorencie2009}Laflorencie, N., Mila, F.: Theory of the Field-Induced BEC in the Frustrated Spin-1/2 Dimer Compound BaCuSi$_2$O$_6$. Phys.Rev.Lett. \textbf{102}, 060602 (2009) 

\bibitem{allenspach2020}Allenspach, S., Biffin, A., Stuhr, U. Tucker, G.S.,  Ohira-Kawamura, S., Kofu, M., Voneshen, D.J., Boehm, M., Normand, B., Laflorencie, N., Mila, F.,  R\"{u}egg, Ch.,: Multiple Magnetic Bilayers and Unconventional Criticality without Frustration in BaCuSi$_2$O$_6$. Phys.Rev.Lett. \textbf{124}, 177205 (2020)

\bibitem{okada2016}M. Okada, M., Tanaka, H., Kurita, N., Johmoto, K. Uekusa, H.,  Miyake, A., Tokunaga, M., Nishimoto, S. Nakamura, M., Jaime, M., Radtke, G., Saul, A.: Quasi-two-dimensional Bose-Einstein condensation of spin triplets in the dimerized quantum magnet Ba$_2$CoSi$_2$O$_6$Cl$_2$: Phys. Rev. \textbf{B 94}, 094421 (2016)

\bibitem{kurita2019}Kurita, N., Yamamoto, D., Kanesaka, T., Furukawa, N., Ohira-Kawamura, S., Nakajima, K., Tanaka, H.: Localized Magnetic Excitations in the Fully Frustrated Dimerized Magnet Ba$_2$CoSi$_2$O$_6$Cl$_2$: Phys. Rev. Let. \textbf{123}, 027206 (2019).

\bibitem{puphal2016}Puphal, P., Sheptyakov, D., van Well, N., Postulka, L., Heinmaa, I., Ritter, F., Assmus, W., Wolf, B., Lang, M., Jeschke, H.O., Valenti, R., Stern, R., R\"{u}egg, Ch.,  Krellner, C.: Stabilization of the tetragonal structure in (Ba$_{1-x}$Sr$_x$)CuSi$_2$O$_6$. Phys. Rev. \textbf{B93}, 174121 (2016)

\bibitem{allenspach2021}Allenspach, S., Puphal, P., Link, J., Heinmaa, I.,  Pomjakushina, E., Krellner, C., Lass, J., Tucker, G.S., Niedermayer, C., Imajo, S., Kohama, Y., Kindo, K., Kr\"{a}mer, S., Horvati\'{c}, M., Jaime, M.,  Madsen, A., Mira, A., Laflorencie, N., Mila, F., Normand, B., R\"{u}egg, Ch., Stern, R., Weickert, F.: Revealing three-dimensional quantum criticality by Sr substitution in Han purple. Phys.Rev.Res. \textbf{3}, 023177 (2021)

\bibitem{jaime2022}Jaime, M.,: Crystal Lattice Witness vs Actor Roles in Correlated Electronic Materials. J. Phys. Soc. Jpn. \textbf{91}, 101005 (2022).













\end{thebibliography}
\end{document}